\documentclass[preprint,showpacs,preprintnumbers,amsmath,amssymb,aps]{revtex4}

\usepackage{graphicx}
\usepackage{dcolumn}
\usepackage{bm}

\newcommand{\beq}{\begin{equation}}
\newcommand{\eeq}{\end{equation}}
\newcommand{\beqar}{\begin{eqnarray}}
\newcommand{\eeqar}{\end{eqnarray}}
\newcommand{\bcen}{\begin{center}}
\newcommand{\ecen}{\end{center}}
\newcommand{\bitem}{\begin{itemize}}
\newcommand{\eitem}{\end{itemize}}

\newcommand{\bra}{\left<}
\newcommand{\ket}{\right>}
\newcommand{\wf}{\bm{\Psi}}
\newcommand{\rar}{\rightarrow}

\begin{document}
\preprint{}

\title{Two-Pulse Atomic Coherent Control (2PACC) Spectroscopy
of Eley-Rideal Reactions. An Application of an Atom Laser.}

\author{Solvejg J\o rgensen}
 \email{solvejg@fh.huji.ac.il}
\author{Ronnie Kosloff}%
 \email{ronnie@fh.huji.ac.il}
\affiliation{%
The Fritz Haber Research Center for Molecular Dynamics,\\
Hebrew University, Jerusalem 91904, Israel
}%


\date{\today}

\begin{abstract}

A spectroscopic application of the atom laser is suggested. The spectroscopy termed
2PACC employs the coherent properties of matter-waves from a two pulse
atom laser. These waves are employed to control a gas-surface 
chemical recombination reaction. 
The method is demonstrated for an Eley-Rideal reaction 
of a hydrogen or alkali atom-laser pulse where the surface target is an
adsorbed hydrogen atom. The reaction yields
either a hydrogen or alkali hydride molecule. 
The desorbed gas phase molecular yield and its internal state is 
shown to be controlled by the time and phase delay between two
atom-laser pulses.
The calculation is based on solving the time-dependent Schr\"{o}dinger equation
in a diabatic framework.
The probability of desorption which is the predicted 2PACC signal
has been calculated as a function of the pulse parameters.
\end{abstract}


\maketitle

\section{\label{sec:level1}Introduction}

Controlling the outcome of a  chemical reaction has been the ultimate goal of chemistry.
Coherent control, a new addition to the quest, is based on  
exploitation of quantum interferences of matter-waves to build a constructive 
interference in the desired channel and a destructive interference in all other channels.
To carry out such a task the molecular matter-wave has to exhibit a coherent property.
To date, all experimental applications of coherent control have been based on imprinting 
the coherent properties of a light/optical source onto the matter 
to be controlled\cite{SRice01,RJGordon97}. 
With the experimental realization 
of Bose-Einstein condensation\cite{MHAnderson95,KBDavis95,CCBradley95,CCBradley97},
a new source of coherent matter-waves has become potentially available. The present paper
explores the direct employment of a matter-wave coherent source 
to achieve the goal of coherent control  of a chemical reaction.

The utilization of a matter-wave source depends on the 
experimental ability to direct and shape the matter-waves from a  
Bose-Einstein condensation (BEC). The waves in a BEC trap constitute
a single many-body matter-wave function which is an analogue
of a photon field in an optical cavity.
An output coupler termed  {\em atom laser} transforms the condensate into
a source of either pulsed or continuous coherent 
matter\cite{MOMewes97,BPAnderson98,EWHagley99,IBloch99,YLCoq01,MKohl02,APChikkatur02}.
We propose to employ this matter-wave for a surface mediated chemical reaction
carried out by a two-pulse atomic coherent control (2PACC) spectroscopy.
The present paper elaborates on previous preliminary studies 
(See Refs. \onlinecite{2pacc:1,2pacc:3}).

BEC has been accomplished for a growing list of atoms
$\{$$^{133}{\bf Cs}$ (Ref. \onlinecite{RGrimm02}),
$^{87}{\bf Rb}$ (Ref. \onlinecite{MHAnderson95}),
$^{85}{\bf Rb}$ (Ref. \onlinecite{SLCornish00}),
$^{23}{\bf Na}$ (Ref. \onlinecite{KBDavis95}),
$^{7}{\bf Li}$ (Refs. \onlinecite{CCBradley95,CCBradley97,FSchreck01}),
$^{6}{\bf Li}$ (Ref. \onlinecite{FSchreck01}),
${\bf K}$ (Ref. \onlinecite{Modugno01}),
${\bf H}$ (Ref. \onlinecite{DGFried98}) and 
meta-stable ${\bf He}$ (Refs. \onlinecite{FPDSantos01,ARobert01})$\}$.
All of the above condensates and additional systems to be prepared, are candidates
for a source for a coherently matter-wave.

The principle of interfering optical waves in a molecular framework
has neatly been demonstrated by Scherer et al.\cite{NFScherer90,NFScherer91}.
Using a sequence of two optical pulses, population is transferred
from the ground to the excited electronic state in the iodine molecule.
The first pulse promotes a portion of the wave function to the excited state,
which then evolves in time.
The second pulse, which is phase-locked relative to the first one, 
excites an additional wave packet to the excited state.. 
The intramolecular superposition of the two wave packets 
is subject to quantum interferences, either
constructive or destructive. 
This gives rise to larger or smaller population on the excited state for a given 
time-delay between the pulses. 
The outcome  is then controlled by the optical phase difference between 
the two pulses and the evolution of 
the initial wave packet\cite{NFScherer90,NFScherer91,RKosloff92,ABartana93}.

The objective of the study is to control a bimolecular surface reaction using
a coherent source of matter-waves shaped as two phase-locked pulses.
The target substrate is a stationary atom adsorbed on a surface. 
The outcome of the reaction is a desorbed molecule in the gas phase
composed from one atom from the matter-wave pulse and one surface atom.
The method is therefore  termed  two-pulse atomic coherent control 
(2PACC) spectroscopy. 
The yield of the reaction is controlled by the time delay 
between the two atomic pulses and  the relative phase between them.

The majority of studies in coherent control have
concentrated on unimolecular processes.
An exception is the study by Abrashkevish et al \cite{PBrumer98}
which theoretically considered a 3D-atom-diatom reactive scattering process.
The 2PACC spectroscopy is aimed at the more difficult task of
controlling a  bimolecular reaction described by $A+B \rar C$.

In contrast to a two-pulse photon field the matter waves are dispersive.
This means that the source of the matter-waves, the BEC condensate,
has to be positioned very close to the target surface. Such a device
is realized in the so-called atom chips\cite{MPAJones03,RFolman02,WHansel01}  
or surface micro-traps\cite{AELeanhardt02,HOtt01}.
In the atom chips for example the BEC has been placed a few hundred microns 
above a metal surface\cite{RFolman02}. The additional step required 
to realize the current proposal is a two-pulsed output coupler directing
a coherent pulse of atoms toward the surface.

A coherent optical spectroscopy employed in surface science
is two-photo emission (2PPE) which has served as the inspiration for
the present proposed 2PACC. In 2PPE, two photons from a laser 
are applied to a surface with or without
adsorbates. 
In 2PPE, the first optical pulse excites a surface electron 
to an intermediate level of either an image potential state or
an unoccupied molecular orbital of the adsorbate.
The second photon exploits this excitation to create a measurable outcome.
The method has been used to characterize
the energetics and dynamics of electrons, atoms and molecules on 
surfaces\cite{Giesen85,Fauster95,Review97,Petek02,shumay98,zhu2000,Wolf2000}.
The application of two laser pulses allows for probing  
the nuclear dynamics of the electronic states 
which could not be studied if only one pulse was applied.
Recently, Petek et al\cite{Petek02} have demonstrated that 
applying a sequence of phase-locked optical pulses 
coherently controls the motion of an alkali atom on the metal surface and 
thereby the desorption process.

The bimolecular surface mediated reaction to be controlled by the 2PACC spectroscopy 
is the Eley-Rideal reaction.
The atomic pulses consist of a coherent atomic source of hydrogen or alkali atoms
which collide with a hydrogen atom chemisorbed on a {\bf Cu}(111)-surface. 
The outcome is an ejected molecule in the gas phase of 
either {\bf H}$_2$ or an alkali hydride.
It will be shown that the application of a second atomic pulse induces the 
matter-wave interferences that enable enhancement or suppression of the 
desorbing flux of molecules relative to a single atomic beam experiment.
The quantum interferences are controlled by the time delay and the relative phase
between the two atomic pulses.

Theoretical studies of Eley-Rideal reactions have been
performed by Jackson et al.\cite{Lemoine02,Lemoine01,Persson99,Persson92}
for an incident hydrogen atom beam interacting with various coated surfaces.
They have observed that hot atom processes, 
where the impinging atom becomes trapped onto the surface, play an important role.
The trapped atoms can react with the adsorbate or they can eventually dissipate
enough energy through collisions with the adsorbates to become immobile.
These dissipative forces cause decoherence and therefore will suppress the 2PACC
control. 
In this paper the energy and charge transfer between the surface 
and the adsorbates are not included.

The dynamics of the 2PACC spectroscopy model
is explored by solving the time dependent Schr\"{o}dinger equation
by a Newtonian propagation method \cite{Ashkenazi95}.
The matter-wave pulses are presented by two
phase related Gaussian-like wave-packets.
The atomic and molecular as well as their interactions 
with the surface are described in a diabatic framework.

The methodology of the 2PACC spectroscopy is presented in Section \ref{sec:model}.
In Section \ref{sec:dynamics} the dynamics of 2PACC is presented.
In Section \ref{sec:results} two simulations of 2PACC spectroscopy are studied one with 
a hydrogen source and another with a lithium atom 
laser both impinging on a {\bf Cu}-surface with chemisorbed hydrogen atoms. 
Conclusions and a general outlook are presented in Section \ref{sec:conclusion}.

\section{The model}
\label{sec:model}

The target of control is the  Eley-Rideal reaction  described by
\beq
{\bf Y} + {\bf H}/{\bf Cu}(111) \rar {\bf YH}  + {\bf Cu}(111)\;\;\;.
\eeq
The atom {\bf Y} is from the atom laser source, 
which in the present model consists of either hydrogen or alkali atoms. 
The matter-wave of {\bf Y} is directed to a {\bf Cu}(111)-surface 
with low coverage chemisorbed hydrogen atoms.
When the wave function of {\bf Y} overlaps with that of the adsorbed atom,
interaction is expected, leading to a recombination that forms
the {\bf YH} molecule. If the newly formed molecule has sufficient
energy it will eventually desorb from the surface to the gas phase.
By shaping the wave function of {\bf Y}, the yield of the desorbing
molecules becomes controlled by constructive or destructive interference.
The simplest controlled wave function is obtained  by a sequence of two pulses
where the time delay and relative phase are the control variables.
It is anticipated that atom lasers will be able
to produce such coherent matter-wave pulses
by an output coupler of a coherent source - the Bose-Einstein condensate (BEC).
For example a coherent atomic pulse can be realized by using 
an optical laser source
to transfer the atom from a trapped (BEC) to an untrapped state (the atom laser).
The relative phase between the two optical pulses can be utilized to phase-lock
the two pulses of the atom matter-wave. 
A time delay between the two optical laser determines also the time delay between the 
two pulsed  atomic laser. 
The coherent properties of interest are projected onto this wave function. 
In this study the wave function of an atom laser composed of a single atom 
is represented by two Gaussian wave functions with a well defined time-delay and phase. 

The binding energy for a hydrogen atom on a {\bf Cu}(111)-surface is roughly 2.4 eV.
Since the bond energy of  {\bf H}$_2$ is approximately twice this value, 
the reaction is very exothermic.
For this reason the desorbing molecule is expected to be vibrationally hot 
due to the large exothermicity.
As one proceeds through the list of atoms 
({\bf Y}$=$$\{${\bf H},{\bf Li},{\bf Na},{\bf K},{\bf Rb},{\bf Cs}$\}$) 
the reaction changes from exothermic to endothermic. 
If the {\bf Cu}(111)-surface is replaced with an {\bf Al}(110)-surface 
all the reactions become again exothermic since the binding energy 
of the hydrogen on this surface is only 1.8 eV~\cite{GDoyen96}.

The present modeling of the Eley-Rideal reaction restricts the motion of the two atoms  
to a collinear configuration normal to the surface.
We consider the reaction between an impinging atom of mass $m_y$ 
located at $z_y$ and a target atom of mass $m_h$ located at $z_h$ which is initially 
adsorbed on a flat static surface.
The coordinate set ($z_h,z_y$) is transformed to a new set of coordinates ($r,Z$)
where $r=z_y-z_h$ is the intramolecular distance 
and $Z=(m_hz_h+m_yz_y)/M$ is the center of mass coordinate.
The total and reduced mass of the system are denoted by $M$ and $\mu$, respectively.
Effects arising from phonons and surface corrugations are not included 
in this model. 

\subsection{Two-dimensional potential energy surfaces}
\label{sec:PES}

To understand the dynamics of either a dissociation reaction of a diatomic 
or a recombination reaction of two atomic species on a surface, 
potential energy surfaces (PES) representing the physics
of breaking and forming a chemical bond is required.
A comprehensive overview of the historical development of potential energy surfaces
is given in Ref. \onlinecite{GRDarling95}.

An electronic structure  model for the dissociation/recombination 
of the H$_2$/Cu system has been suggested by
Holloway and co-workers\cite{JHarris88,MRHand89}. Their model is based on 
a small complex of the form {\bf Cu}$_2${\bf H}$_2$ in a planar $C_{2v}$ geometry. 
The intramolecular coordinates are  $r$ which is the distance between the two hydrogen atoms,
and $Z$ the distance between the midpoint of {\bf H}-{\bf H} and {\bf Cu}-{\bf Cu}.
For large value of $Z$ there is no interaction between
the two units, {\bf H}-{\bf H} and {\bf Cu}-{\bf Cu}, meaning that
each unit has a separate set of fully occupied molecular orbitals.  
As $Z$ decreases the two units approach each other,
the orbitals with similar symmetry with respect to the bond 
center mix, e.g. the parallel-bonding orbital of {\bf H}-{\bf H} and {\bf Cu}-{\bf Cu}
interfere with the parallel anti-bonding of {\bf Cu}-{\bf H}.
Due to these interactions the orbital energies shift.
This causes a crossing between the "parallel" and "perpendicular" orbitals with the result that
the {\bf H}-{\bf Cu} bonding becomes energetically favorable.

The PES of the reactant surface has been constructed from a Morse potential describing 
the hydrogen-hydrogen bond  and a repulsive potential  as 
the hydrogen molecule approaches the surface.
In the product channel the two separated hydrogen atoms are chemisorbed on the surface,
this bond is described by a Morse potential.
The hydrogen-hydrogen repulsion on the surface is represented by 
an exponential repulsive potential.

In the diabatic representation, the potential is represented by a 2$\times$2 ${\bm V}$-matrix
\beq
{\bm V}(r,Z)= 
           \left [ \begin{array} {c}
                   {\bm V}_{RR}(r,Z)\;\;\; {\bm V}_{RP}(r,Z) \\
                   {\bm V}_{PR}(r,Z)\;\;\; {\bm V}_{PP}(r,Z)
                              \end{array}
                     \right ] \;\;\;.
\eeq
The diagonal elements are the potential energy surfaces of the reactant and the product 
states and the off-diagonal elements is the non-adiabatic couplings between them.
Upon diagonalization of the 2$\times$2 ${\bm V}$-matrix two new adiabatic PES
are obtained.
In the previous studies of Eley-Rideal reactions the adiabatic PES has been used.
We will briefly discuss the idea behind them.

Jackson et al.\cite{Lemoine02,Persson99,Hammer99} used the London-Eyring-Polanyi-Sato (LEPS)
potential energy surface for modeling the Eley-Rideal reactions.
The LEPS potential is given by
\beq
V= U_m+U_a+U_b \pm \sqrt{A_m^2+[A_a+A_b]^2-A_m[A_a-A_b]}\;\;,
\eeq
where $U$ and $A$ are the Coulomb and exchange contributions.
The intramolecular contributions are denoted $U_m$ and $A_m$, whereas the other terms
describe the atomic interactions with the surface. 
Here, only the lowest adiabatic PES are considered.
The PES of the molecule as well as the interactions between 
the surface and a single atom have been calculated by Density Functional Theory (DFT).
They have been fitted to the functional forms, $U$ and $A$, 
which are based on Morse-like attractive and repulsive terms
which decay exponentially with increasing distances.
Surface effects have been introduced through expanding the Morse parameters 
representing the surface in a Fourier series of the reciprocal lattice vectors. 

The LEPS-potential is a member of a class of diatomic in molecules (DIM) 
potential forms. For surface reactions, an extension called 
embedded diatomic in molecules (EDIM) has been developed.
The model was originally developed by Truong et al \cite{TNTruong89}
and recently used for the description of the motion of $N_2$
on a ruthenium-surface\cite{CTully02}.
In the EDIM the intramolecular interactions are modeled by Morse and
anti-Morse potentials which represent the singlet and triplet electronic state
of the diatomic molecule in the gas phase.
The interaction between the atoms in the gas phase and the surface
is modeled by embedded atom model (EAM), in which the atom experiences 
an average charge density from the surface. 

In the 2PACC spectroscopy the coherent properties of the encounter is
intricate therefore 
the Born-Oppenheimer approximation
in which the reaction takes place on a single potential energy surface has to be replaced
with a non-adiabatic framework.
In the 2PACC dynamics 
a diabatic frame is chosen consisting of two potential energy surfaces.
Specifically, for the Eley-Rideal reaction the diabatic PES are constructed from
two atomic or one molecular chemical species interacting with the surface. 
In the reactant channel the interaction between the stationary hydrogen atom 
and the surface is strong due to a chemical bond.
We are using a semi-empirical functional form for the PES.
The impinging atom experiences a repulsive force from the adsorbed atom as
well as a long-range attraction from the surface due to polarization forces.
In the product channel the newly formed molecule is attached to the surface 
by a weak bond induced by the polarization forces between the surface and the molecule.
The PES for the reactant and product channels have 
the following functional forms 
\beqar
{\bm V}_{RR}(r,Z) &=& V^R_{HM}+V^R_{YM}+V^R_{YH}
\label{eq:reactant_pes}\\
{\bm V}_{PP}(r,Z) &=& V^P_{YH-M}+V^P_{YH}\;\;,
\label{eq:product_pes}
\eeqar
where the index {\bf M} represents the interactions with the surface.
The last terms of Eqs.(\ref{eq:reactant_pes}-\ref{eq:product_pes})
represent the intramolecular interaction between the two gas phase atoms, {\bf H} and {\bf Y}. 
The other terms represent the interaction between an atom or a molecule with the 
surface. 
In the following section we will discuss each 
of the individual terms in the reactant and product PES.

\begin{table*}[tbh!]
\caption{
\label{table:parameter}
The parameters for PES for the Eley-Rideal reaction on a {\bf Cu}(111)-surface.
The hydrogen atom is chemisorbed on a hollow site of the surface.
}
\begin{ruledtabular}
\begin{tabular}{lccc}
hydrogen molecule  & $D_{HH}$$=$4.505 eV\footnotemark[1]       & $Z_e$$=$2.0{\AA}            & $\Delta=0.2$\\
                   & $r_{e}^{ad}$$=$0.741{\AA}\footnotemark[1] & $r_{e}^g$$=$0.754{\AA}\footnotemark[2]  &             \\
                   & $\alpha_{HH}^g$$=$2.2{\AA}$^{-1}$  & $\alpha_{HH}^{ad}$$=$2.11{\AA}$^{-1}$ &        \\\hline
                  & & & \\
chemisorption     & $D_{HM}$$=$2.334eV\footnotemark[1] & $z_{HM}^e$$=$0.916{\AA}\footnotemark[1]  & $\alpha_{HM}$$=$1.75{\AA}$^{-1}$ \footnotemark[1] \\
                  & & & \\\hline
Physisorption& $A_M$/eV              & $b_M$/{\AA}$^{-1}$        & $C_M$/eV{\AA}$^{-3}$ \\ 
{\bf H}$_2$ or {\bf H}   &  600                   & 3.8                       &  3.5           \\
{\bf LiH} or {\bf Li}   &  650                   & 3.15                      &  7.0           \\
{\bf NaH} or {\bf Na}   &  760                   & 2.7                       &  12.5          \\
{\bf KH} or {\bf K}     &  850                   & 2.4                       &  20.0          \\
{\bf RbH} or {\bf Rb}   &  930                   & 2.23                      &  27.0          \\
{\bf CsH} or {\bf Cs}   &  950                   & 2.1                       &  34.0          \\
\end{tabular}
\end{ruledtabular}
\footnotetext[1]{Ref.\onlinecite{Hammer99}} 
\footnotetext[2]{Ref.\onlinecite{WKolos65}}
\end{table*}

\subsubsection{Intramolecular interactions}
Asymptotically, the intramolecular interaction potentials, 
$V^R_{YH}$ and $V^P_{YH}$ in the reactant and product channels
become the triplet and singlet electronic states
of the molecule, {\bf YH}, when both atoms are far from the metal surface.
The PES of the alkali hydride molecule  was based on the intramolecular 
gas-phase singlet and triplet electronic states 
which have been evaluated by a multi-configuration self-consistent field calculation
by Geum et al.\cite{NGeum01}.
These potentials have been fitted to a cubic spline  interpolation function.

For the hydrogen molecule the singlet and triplet PES 
are given by a Morse and an anti-Morse potential
\begin{eqnarray*}
V^P_{HH}(r,Z) &=& D_{HH}([1-e^{\{-\alpha_{HH}(Z) (r-r_{HH}^e(Z))\}}]^2-1)
\label{eq:morse}\\
V^R_{HH}(r,Z) &=& \frac{1}{2} \; \frac{1-\Delta}{1+\Delta}\;
              D_{HH}([1+e^{\{-\alpha_{HH}(Z) (r-r_{HH}^e(Z))\}}]^2-1) \;\;.
\label{eq:antimorse}
\end{eqnarray*}
The dissociation energy of the hydrogen molecule is denoted $D_{HH}$ 
and $(1-\Delta)(1+\Delta)$ is the Sato factor.
The equilibrium distance between the two hydrogen atoms, $r_{HH}^e$,
and the coupling strength, $\alpha_{HH}$,
have been obtained as a function of $Z$ by an interpolation
between values for the gas phase to  ones for the adsorbate:
\beqar
\alpha_{HH}(Z) & = &\alpha_{HH}^{ad}  \;\;\;\;\;\;\;\;\;\;\;\;\;\;\;\;\;\;\;\;
\;\;\;\;\;\;\;\;\;\;\;\;\;
\;\;\;\;\;\;\;\;\;\;\;\;\;\;\;\;\;\;\;\; for \;\;\;\;\ Z \le Z_e
\nonumber\\
\alpha_{HH}(Z) & =& \alpha_{HH}^{g}-
                (\alpha_{HH}^{g}-\alpha_{HH}^{ad})\Gamma_4(Z,Z_e,2b_{HM})
                \;\;\;\; for \;\;\;\; Z>Z_e
\eeqar
and
\beq
r_{HH}^e(Z)= r_e^g-(r_e^g-r_e^{ad}) \Gamma_4(Z,Z_e,2b_{HM}) \;\;.
\eeq
The superscripts $g$ and $ad$ indicate the gas phase and the adsorbed
state.
Table \ref{table:parameter} gives the parameters used for the PES for a {\bf Cu}(111)-surface
covered by a hydrogen atom adsorbed on the hollow site.

\begin{figure}[tb!]
{\rotatebox{270} {\includegraphics[scale=0.4]{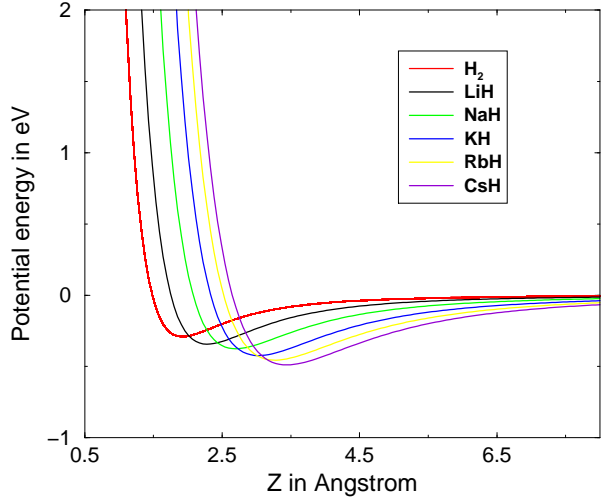}}}
\caption{\label{fig:phys} The potential energy surface representing the physisorption 
of a hydrogen or an alkali hydride molecule on a {\bf Cu}(111)-surface are shown as a function of the distance between the surface and the mass center of the diatomic.}
\end{figure}

\subsubsection{Surface interactions}

In the product channel a molecule ({\bf YH}) is formed and eventually 
transferred to the gas phase. 
The molecule is coupled to the surface by a weak bond 
induced by the polarization forces.
The interaction potential, $V^P_{YH-M}$,
is represented as a sum of two terms - a short range
repulsive term and a weak long-range attraction 
between the molecule and the surface
\beq
V^P_{YH-M} =  A_{M} e^{(-b_{M} Z)} - 
              \frac{C_{M}}{Z^3}(1-\Gamma_4(Z,0,2b_{M}))\;\;.
\label{eq:vphys}
\eeq
The incomplete Gamma function which turns off the attraction as the
{\bf Y}-atom approaches the surface is given by:
\beq
\Gamma_m(x,x_0,a)=\sum_{k=0}^{k=m} 
        \frac{a(x-x_0)^k}{k!} e^{(-a(x-x_0))}\;\;.
\label{eq:gamma}
\eeq
The parameterization of the physisorption of the molecular interaction with
the surface has been chosen such that the minimum energy e.g. 
the physisorption energy is observed at $Z_e$$=$$(r_{HM}^e+m_y r_{YH}^e)/M$. 
The location of the minimum energy is moved further away from the surface 
as the mass of the impinging atom increases.
As {\bf Y} becomes heavier, the polarization force between the surface 
and the alkali atom increases, thereby the binding energy 
of the physisorption bond becomes stronger. 
The amplitudes of the repulsion, $A_{M}$ and attraction, $C_{M}$
increase as  one proceeds down the list of atoms, 
$\{${\bf H},{\bf Li},{\bf Na},{\bf K},{\bf Rb},{\bf Cs}$\}$.
The coupling strength, $b_{M}$, is inversely proportional to $Z_e$.
The parameters are constructed so that the physisorption energy of {\bf YH} 
increases slightly as one proceeds down the list of atoms 
$\{${\bf H},{\bf Li},{\bf Na},{\bf K},{\bf Rb},{\bf Cs}$\}$.
The physisorption potentials are shown in Figure \ref{fig:phys}
and the parameters are given in Table \ref{table:parameter}.

To model the interaction between the impinging atom in the reactant channel 
a physisorption potential has been used. 
The same set of parameters for the atomic interaction as for the molecular
physisorption interactions have been chosen 
but the coordinate, $Z$, in Eq.(\ref{eq:vphys}) is replaced by 
the location of the impinging atom, $z_y$.

In the reactant channel the interaction between the chemisorbed atom and the surface is described by
a Morse potential
\beq
V^R_{HM} = D_{HM}([1-e^{(-\alpha_{HM} (z_h-z_{HM}^e))}]^2-1) \;\;,
\eeq
where $D_{HM}$ is the dissociation energy of the atom-metal bond and 
$z_{HM}^e$ is the equilibrium bond length. 
The parameters are given in Table \ref{table:parameter} for 
a hydrogen atom adsorbed at the hollow site of a {\bf Cu}(111)-surface.

\subsection{Wave function of the atomic pulse}

Initially, when the impinging atom is far away from the surface and the
adsorbed atom, the wave function representing the outcome of the total system is written as 
a product of a wave function on each atom,
\beq
\Psi_{atom}(r,Z)=N \psi_h(z_h,z_y=z_s)\;\psi_y(z_y)\;e^{ik_yz_y}
\;\;;\;\;\psi_y(z_y)=e^{-(z_y-z_s)^2/\sigma_y}
\label{eq:atomlaser}
\eeq
where $N$ is a normalization factor.
The wave function of the adsorbed atom, $\psi_h$, was chosen as
the lowest energy eigenstate
of the PES of the reactant channel along the one-dimensional path 
with a fixed distance, $z_s$, between the approaching atom and the surface.
Considering only the lowest vibrational states would be sufficient since in 
a given experimental situation one would expect that the temperature of the surface
is ultracold.
The wave function of the approaching atom, $\psi_y$, is represented by
a Gaussian wave function centered at $z_y=z_s$ and with a variance, $\sigma_y$.
The momentum of the approaching atom is denoted by $k_y=\pm \sqrt{2m_y E /\hbar^2}$ 
which is related to the energy $E$ of the propagating atom laser.
The sign of the momentum operator determines the orientation of the atom 
laser: For negative momentum, the atom moves toward the surface.
The variance is related to the dispersion of the atom laser.

\section{2PACC Dynamics}
\label{sec:dynamics}

The 2PACC spectroscopy of an Eley-Rideal reaction consists of the dynamics on
the coupled reactant and product potential energy surfaces.
The wave function of the system is described by the vector 
\beq
\wf(r,Z,t)=\left ( \begin{array} {c}
                   \wf_{R}(r,Z,t) \\
                   \wf_{P}(r,Z,t)
                              \end{array}
                     \right ) \;\;\;,
\eeq
where the wave functions in the reactant and product channels 
are denoted by the index $\{R,P\}$. 
The Hamiltonian of the system is represented by 
\beqar
{\bm H}(r,Z)& =& {\bm T}(r,Z) + {\bm V}(r,Z) \nonumber\\
          & =&  \left [ \begin{array} {c}
                   {\bm T}_{RR}(r,Z) \;\;\;\;\;\;\;\;\;\;\;\;\;{\bm 0}  \\
                   \;\;\;\; {\bm 0} \;\;\;\;\;\;\;\;\;\;\;\;\;   {\bm T}_{PP}(r,Z)\;\;\;
                              \end{array}
                     \right ] 
                   +
           \left [ \begin{array} {c}
                   {\bm V}_{RR}(r,Z)\;\;\; {\bm V}_{RP}(r,Z) \\
                   {\bm V}_{PR}(r,Z)\;\;\; {\bm V}_{PP}(r,Z)
                              \end{array}
                     \right ] \;\;\;.
\eeqar
The diagonal elements of the Hamiltonian have the following form
\beq
{\bm H}_{ii} (r,Z) =  {\bm T}_{ii} (r,Z)+ {\bm V}_{ii} (r,Z)= -\frac{\hbar^2}{2M}\frac{\partial^2}{\partial Z^2}
                     -\frac{\hbar^2}{2 \mu}\frac{\partial^2}{\partial r^2}
                     +{\bm V}_{ii}(r,Z) \;\;\;\;\;\;\;\;\;i=\{R,P\}
\eeq
where the first two terms are the kinetic energy operator 
for the two degrees of freedoms ($r$ and $Z$) and the last term is 
the potential energy function
which was described in Section \ref{sec:PES}.
The off-diagonal elements of the Hamiltonian represent the non-adiabatic 
coupling between the two channels which are described by
\beq
{\bm V}_{RP}(r,Z) = {\bm V}_{PR}(r,Z) = e^{-(r-r_{YH}^e)^2}\;e^{-\beta Z}\;\;,
\eeq
where $\beta$ is the non-adiabatic coupling strength and $r_{YH}^e$ is the equilibrium distance
between the two atoms in the singlet molecular PES.
This representation ensures that the electron density of the metal decays 
exponentially outside the surface into the vacuum. 
It is important to note that the adiabatic PES of the  Eley-Rideal reaction
which have been used by Jackson et al.\cite{Lemoine02,Lemoine01,Persson99,Persson92}
can be obtained by diagonalization of this 2$\times$2 ${\bm V}$-matrix.

The dynamics of the 2PACC was followed by solving the
time-dependent two-channel Schr\"odinger equation which is given by
\beq
i \hbar \frac{\partial \wf }{\partial t} = {\bm H} \wf  \;\;\;.
\label{eq:timeSchr}
\eeq

In the two-pulse atomic coherent control (2PACC) spectroscopy
the first atom laser represented by the wave packet (Eq.(\ref{eq:atomlaser})) 
is initialized at the reactant PES at a time $t=0$ e.g 
\beq
\wf(r,Z,0)=\left ( \begin{array} {c}
                   \Psi_{atom} (r,Z) \\
                   0
                              \end{array}
                     \right ) \;\;\;.
\eeq
This initial wave packet evolves in time and 
after a time delay, $\Delta t$, the second atom pulse is introduced given by
\beq
\Psi_{atom,2}(r,Z) = \Psi_{atom}(r,Z) \exp{(-i\theta)}
\eeq
where $\theta\in[-\pi;\pi]$ is the relative phase between the two atomic pulses
that describes the coherence between them.
As the wave packet propagates population is transferred from the reactant 
to the product PES through the non-adiabatic coupling.

\begin{table}
\caption{
\label{table:nummerical}
Computational parameters for the wave packet propagation of the 2PACC dynamics.}
\begin{ruledtabular}
\begin{tabular}{ll}
time step               &     $\delta t$$=$0.097 fs                \\
propagation time        &     $t_{max}$$=$484 fs                   \\
time steps              &     $N_t$$=$5000                         \\
                        &                                          \\
grid points along $r$   &     $N_r$$=$256                          \\
grid spacing along $r$  &     $\Delta r$$=$0.0529 {\AA}            \\
grid starts at          &     $r_{min}$$=$0.0529 {\AA}             \\
grid points along $Z$   &     $N_Z$$=$256                          \\
grid spacing along $Z$  &     $\Delta Z$$=$0.0529 {\AA}            \\
grid starts at          &     $Z_{min}$$=$0.0529 {\AA}             \\
                        &                                          \\
variance                &     $\sigma_y$$=$0.280 {\AA}$^{2}$       \\
initial position        &     $z_s$$=$6.82 {\AA}                   \\
momentum                &     $k$$=$9.45 {\AA}$^{-1}$              \\
                        &                                          \\
absorbing potential     &     $\Delta_r$$=$$\Delta_Z$$=$1.32 {\AA} \\
                        &     $V_0$$=$0.00027 eV                   \\
                        &                                          \\
non-adiabatic coupling   &     $\beta$$=$0.027 eV                   \\
dividing flux line      &     $Z_{flux}$$=$5.24 {\AA}              \\
\end{tabular}
\end{ruledtabular}
\end{table}

\subsection{Computational Method}
\label{sec:computational}
The wave function is represented on a two-dimensional grid.
First, the wave function of the chemisorbed hydrogen was calculated.
Using an imaginary time propagation\cite{Kosloff86} the one-dimensional wave function of
the hydrogen atom, $\psi_H$ along the line with a fixed distance
between the approaching atom and the surface, $z_y$$=$6.8{\AA}
has been relaxed to its vibrational ground state.
Representing the wave function of the impinging atom as a Gaussian wave function,
the total wave packet Eq.(\ref{eq:atomlaser}) is then initialized.

The dynamics of the 2PACC is obtained by propagating the initial wave function
by $\exp{(-i{\bm H}t)}\psi(0)$,
in which the time-evolution operator $\exp{(-i{\bm H}t)}$
is expanded by Newtonian interpolation polynomials 
with Chebychev sampling points\cite{Kosloff94b}.
The kinetic energy operator has been evaluated using the fast Fourier 
transformation technique\cite{Kosloff88,Kosloff94b}.
The parameters used in the wave packet propagation are displayed 
in Table \ref{table:nummerical}.

The scattered wave function is removed at large values of $r$ and $Z$ 
by complex absorbing potentials\cite{Vibok92a} which prevent 
reflection and transmission at the end of the grid.
The overall potential can be written as
\beqar
V(r,Z)  & = & V(r,Z) + V_{abs}(r) + V_{abs}(Z)
                             \;\;\;\;if\;\;\;\;     r-\Delta_r \le r \le r_{max}
                             \;\;\;\; and \;\;\;\;   Z-\Delta_Z \le Z \le Z_{max}
\nonumber\\
       & =  & V(r,Z) + V_{abs}(r)  \;\;\;\;\;\;\; \;\;\;\;\;\;\;
                             \;\;\;\;if\;\;\;\;  r-\Delta_r \le r \le r_{max}
                             \;\;\;\; and \;\;\;\; Z\le  Z-\Delta_Z
\nonumber\\
       & =  & V(r,Z) + V_{abs}(Z)  \;\;\;\;\;\;\; \;\;\;\;\;\;\;
                             \;\;\;\;if\;\;\;\; r \le  r-\Delta_r
                             \;\;\;\; and \;\;\;\;   Z-\Delta_Z \le Z \le Z_{max}
\nonumber\\
       & =  & V(r,Z)         \;\;\;\;\;\;\;\;\;\;\;\;\;\;\;\;\;\;\;\;\;\;\;\;\;\;
                             \;\;\;\;if\;\;\;\; r\le  r-\Delta_r
                             \;\;\;\; and \;\;\;\; Z\le  Z-\Delta_Z
\eeqar
where 
the complex potential is given by
$V_{abs}(r)=iv_0(r-(r_{max}-\Delta_r))^2$.
The same functional form is used for $Z$.
$r_{max}$ is the last grid point and $\Delta_r$ is the interval where
the complex potential is applied. 

As the wave function is evolving from the entry/reactant channel,
the non-adiabatic coupling term is responsible for transferring the 
amplitude between the two diabatic surfaces. 
Eventually the molecule desorbs
from the metal surface.
This leads to an outgoing flux in the exit/product channel.
The probability of a {\bf YH}-molecule to escape from the metal surface is obtained via a 
flux-resolved analysis carried out at an asymptotic value of 
$Z$$=$$Z_{flux}$$=$5.24{\AA}.
The total accumulated desorbing flux which is the predicted 2PACC signal is computed by
\beqar
F & = & \sum_i^{N_t} J(t_i) \delta t\\
  & = & \frac{\delta t}{M} \sum_i^{N_t} Im 
            \Big [~ \int dr \;
                 \wf_P^*(r,Z_{flux},t_i) 
                  \frac{ \partial \wf_P(r,Z,t_i)}{\partial Z}
            \Big |_{Z=Z_{flux}}
            \; \Big ]\;\;,
\nonumber
\eeqar
where the derivative is evaluated with a Fourier Transform.
The integrated flux has been determined as a function of the time delay, $\Delta t$,
and the phase-relation, $\theta$, between the two atom laser pulses.

Furthermore, the accumulated flux current can be evaluated 
for each of the vibrational states along the dividing line
\beqar
P_n & =& \sum_i^{N_t} j_n(t_i) \delta t\\
    & = & \frac{\delta t}{M} \sum_i^{N_t} Im
            \Big [ \;
                  \wf_n^*(Z_{flux},t)
                  \frac{\partial \wf_n(Z,t)}{\partial Z}
                  \Big |_{Z=Z_{flux}}
            \; \Big ] \;\;,
\nonumber
\eeqar
where $j_n$ is the probability current for the wave packet to go 
into the n'th vibrational state.
\begin{figure}[htb!]
{\includegraphics[scale=0.5]{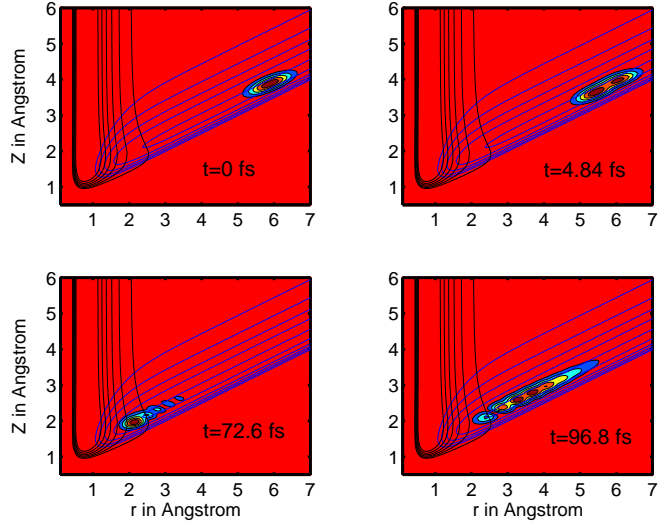}}
\caption{
\label{fig:wavepacket_reactant}
The evolution of the wave  packet ($|\wf_R|^2$) from an  atom laser source applied
to a {\bf Cu}(111)-surface coated with hydrogen atoms superimposed  on the
reactant potential energy surface. Snapshots are shown
for a time delay of $\Delta t$$=$4.84 fs and the phase, $\theta$$=$$-\frac{1}{2} \pi$.
The contour values of potential energy surfaces are  -3, -2.5, -2, -1.5, -1 and -0.5 eV.
The PES of the reactant and
the product channels are shown with blue and black lines, respectively.
}
\end{figure}
Here, $\wf_n(Z,t)$  is the projection of the wave function onto the vibrational eigenstates,
$\chi_n$, along the dividing line
\beq
\wf_n(Z,t) = \int dr \chi^*_n (r) \wf_P(r,Z,t) \;\;.
\eeq
The vibrational eigenstatates have been calculated by imaginary time propagation.
(See Appendix \ref{app:a} for further details).

\begin{figure}[!tb]
{\includegraphics[scale=0.5]{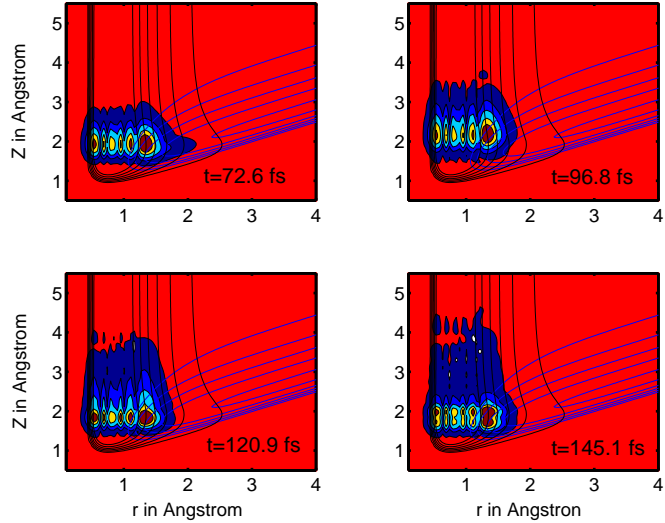}}
\caption{
\label{fig:wavepacket_product}
The evolution of the wave packet on the product channel ($|\wf_P|^2$) 
are shown for time delay of $\Delta t$$=$4.84 fs and phase, $\theta$$=$$-\frac{1}{2}\pi$.
Notice the exciting wave packet on the reactant channel at 145.1 fs.
Parameters used for this calculation are given in Figure \ref{fig:wavepacket_reactant}.
}
\end{figure}

\begin{figure}
{\includegraphics[scale=0.5]{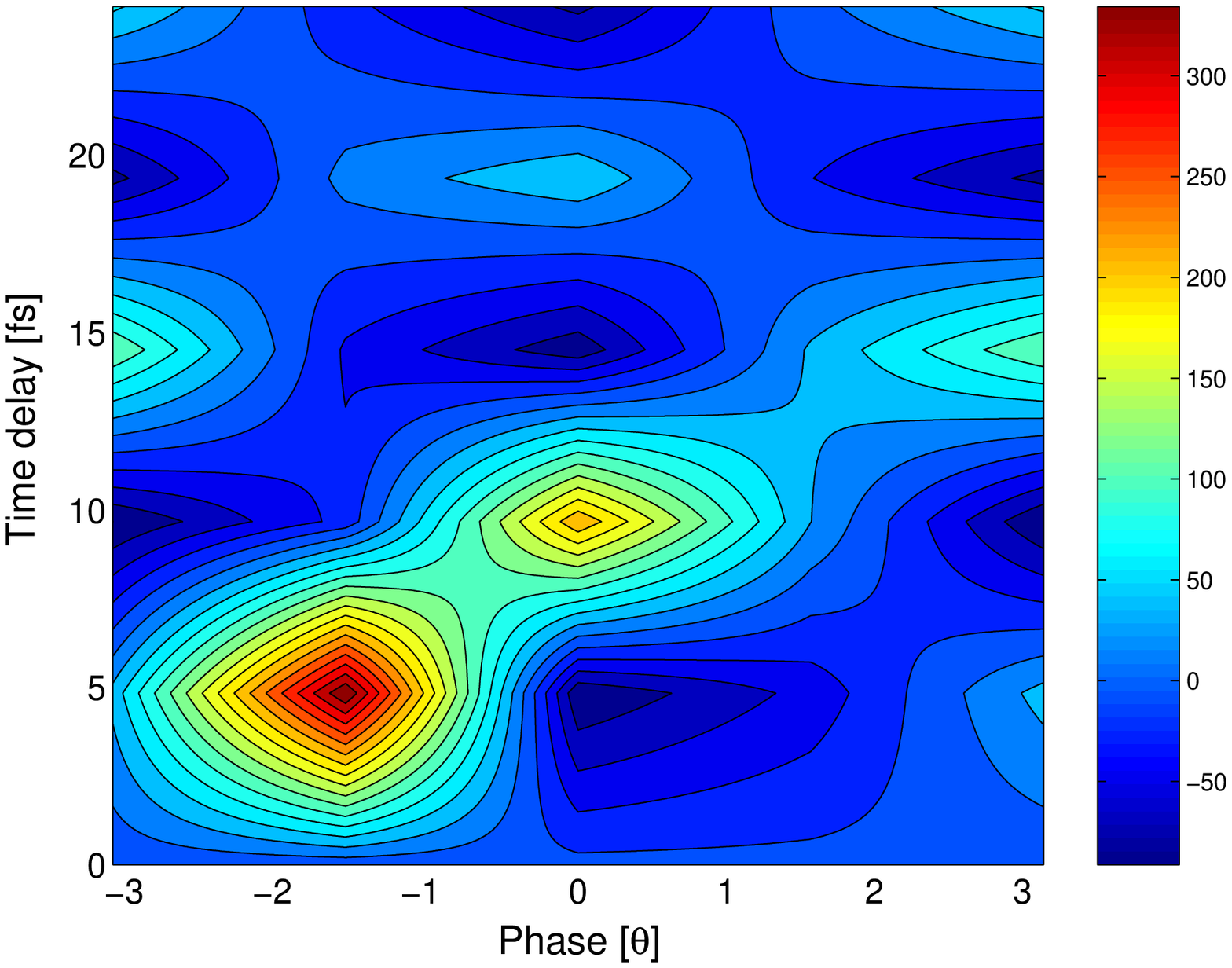}}
\caption{
\label{fig:2PACC}
The 2PACC signal of a hydrogen atom laser impinging on a surface
with chemisorbed hydrogen atoms relative to the outcome for one pulse (in $\%$).
The signal is  shown as a function of the time delay
(in fs) and relative phase ($\theta$)
between the two atomic pulses for an initial wave packet with energy -2.3 eV.
The predicted 2PACC signal is calculated from
the integrated flux along a dividing line on the
product channel at $t=484$ fs.
Areas of enhancement are in red and those of depletion in blue.
}
\end{figure}

\section{Results and Discussion}
\label{sec:results}

The 2PACC of a hydrogen source is compared to that of a lithium atom laser source
impinging on a surface with chemisorbed hydrogen atoms.
The evolution of the wave packet on the reactant and the product channel are presented.
The 2PACC signal and the vibrational analysis of the desorbing molecule are
displayed.  
The control parameters are  the time delay and relative phase between
the two atom laser pulses.

\subsection{H+H/{\bf Cu}(111)}
The dynamics of the 2PACC spectroscopy are demonstrated for a hydrogen
atom laser source impinging on a {\bf Cu}(111)-surface coated with hydrogen atoms.
Compared to earlier calculations\cite{2pacc:1,2pacc:3} the grid spacing
in both degrees of freedom ($r$,$Z$) was reduced by a factor of 2. 
The initial starting position of the wave function was $z_s$$=$6.8{\AA} 
compared to 13.6{\AA} in the earlier calculations.
At the shorter distance the interactions between the impinging atom 
and the adsorbate and the surface is still negligible.

In Figures \ref{fig:wavepacket_reactant} and \ref{fig:wavepacket_product}
snapshots of the evolution of the wave packet on the reactant and the 
product surfaces are shown for 
$\theta$$=$$-\frac{1}{2}\pi$ and $\Delta t$$=$4.84fs. 
First, the initial wave packet (Eq.(\ref{eq:atomlaser})) is generated and 
this atom laser pulse evolves in time. 
The energy of the initial wave packet is -2.3 eV and the impinging atom 
has a kinetic energy of 0.2 eV.
After the specified time delay a wave packet - the second atom laser pulse -
is placed in the original position of the first wave-packet
(see second snapshot of Figure \ref{fig:wavepacket_reactant}).
The relative phase between the two wave packets at the initial position
may be different e.g. $\theta \neq 0$.
The wave function which resembles the two atom laser pulses propagates toward
the non-adiabatic region where the reactant and the product surfaces intersect.
As the wave packet enters the non-adiabatic region 
a part of the wave function is transferred to the molecular state where it can exit.
This means that the approaching hydrogen atom has
reacted with the adsorbed hydrogen and a H$_2$ molecule is formed.
The non-reactive part of the wave function collides with the adsorbed hydrogen 
atoms. 
Either it scatters back to the gas phase or it gets trapped on the surface
due to the weak polarization forces between the impinging atom and the surface (physisorption).

\begin{figure}[tb!]
{\includegraphics[scale=0.5]{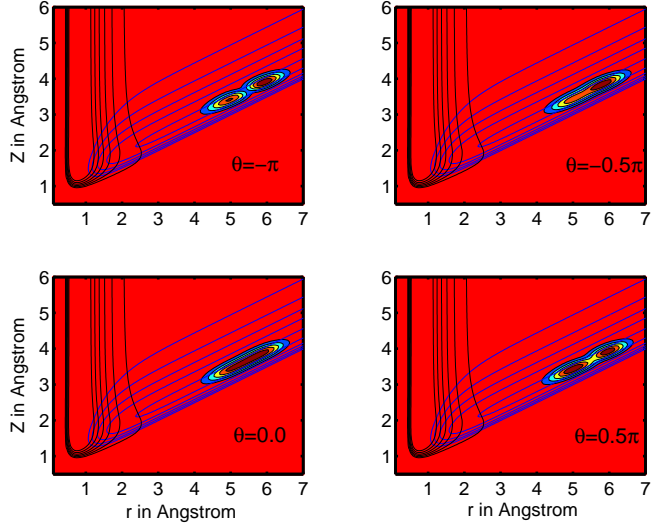}}
\caption{
\label{fig:2wp}
The wave packet after the second atom laser pulse has been applied at $t$$=$$\Delta t$$=$14.5 fs
for different relative phases between the two atomic pulses.
}
\end{figure}

The population which is transferred through the non-adiabatic coupling to the 
product channel builds up slowly as a function of time. 
In the beginning the newly formed hydrogen molecule is physisorbed on the surface. 
That is, it is trapped in the potential well. 
Later a fraction of the wave function exits the channel and the hydrogen molecule 
desorbs from the surface to  be detected in the gas phase. 
The desorbing molecule is vibrationally excited. 
The node structure of the exiting wave function shows that the hydrogen molecule
desorbs in the fifth vibrational state ($\nu$$=$4). 
This observation will be verified by a vibrational analysis.

\begin{figure}[tb!]
\includegraphics[scale=0.5]{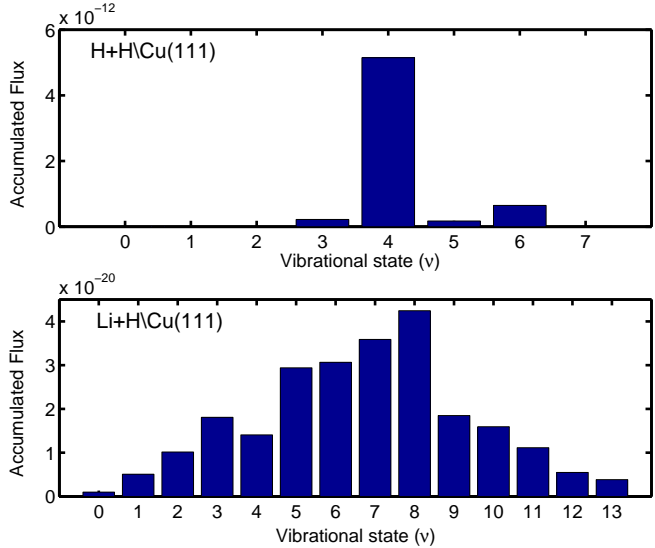}
\caption{\label{fig:vibone} 
The accumulated flux in each of the vibrational states for the one pulse experiment are shown for
hydrogen (upper figure) and  lithium (lower figure).  }
\end{figure} 
The 2PACC signal, which is the difference between a one pulse
and a two pulse desorbed molecular outcome is calculated from
the total integrated flux at $t=484$ fs.
Figure \ref{fig:2PACC} shows the 2PACC signal as a function of time delay and relative
phase between the two pulses.
The amount of control of the 2PACC signal demonstrated in this model is
an enhancement of 350 $\%$ for $\theta$$=$-$\frac{1}{2}\pi$ and $\Delta t$$=$4.84 fs
relative to one atomic pulse compared with a suppression of 95\%
at $\theta$$=$$\pi$ and $\Delta t$$=$9.7 fs. 

The  2PACC signal shows a variation with respect
to both control parameters, meaning that the outcome of
the Eley-Rideal reaction is coherently controlled by the time delay and the relative phase
between the two atomic pulses.
The application of the second pulse creates a quantum interference with the first pulse.
Such interferences are either constructive or destructive resulting in
increasing or decreasing the flux of the desorbing molecules in the product channel.
The effect of these interferences  can be visualized by considering the wave function just after 
the second atom pulse has been applied.
In Figure \ref{fig:2wp} the total wave packet is shown for $t$$=$$\Delta t$$=$14.5fs for
different relative phases between the two atom laser pulses.
An elongated structure of the wave packet immediately after the second
pulse gives rise to destructive interference and decrease of the desorption yield whereas a
"node-like" structure gives rise to constructive interference and an enhancement of the 
yield.

\begin{figure*}[tb!]
\includegraphics[scale=0.9]{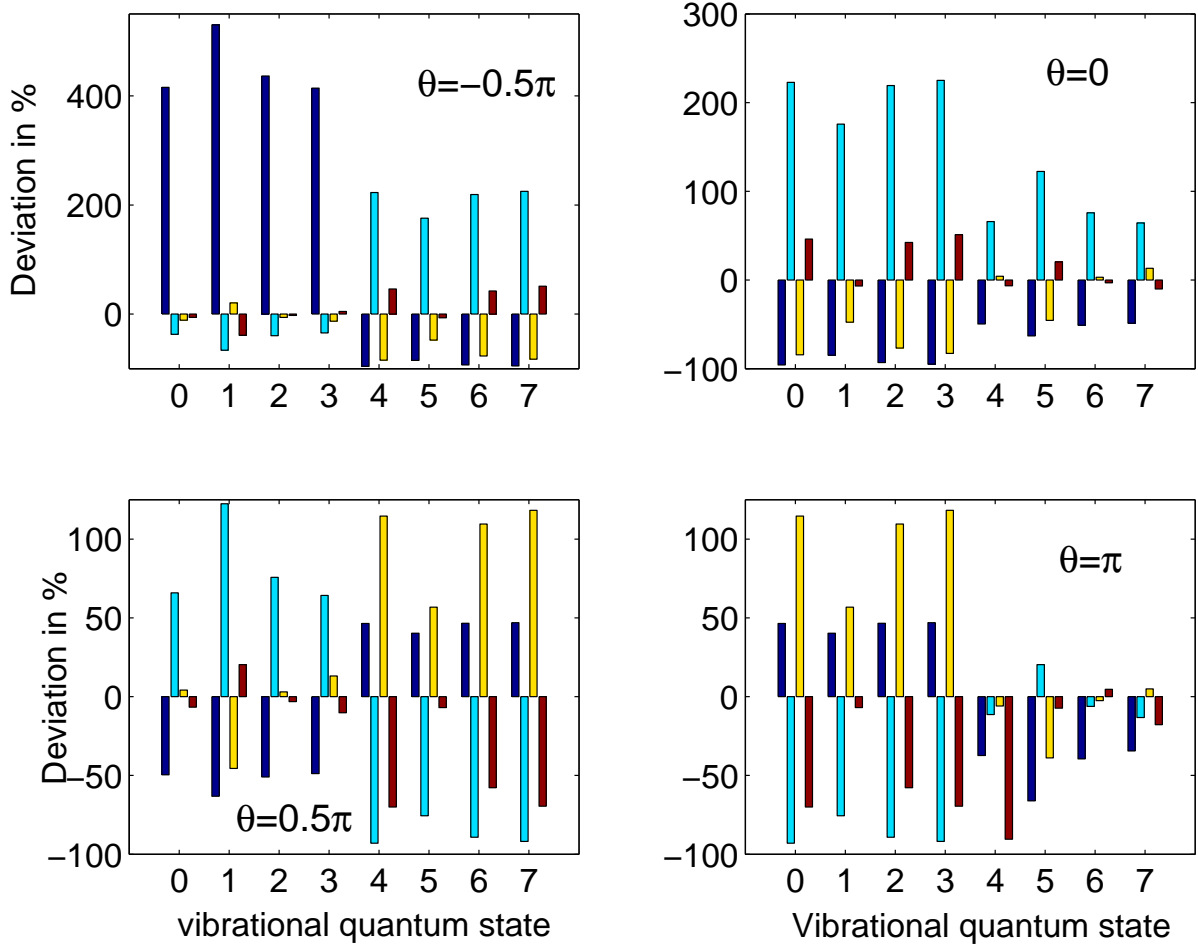}
\caption{\label{fig:hhvib1}
The percentage deviation in the accumulated flux in the vibrational states
from the one pulse experiment are shown for four different
relative phases ($\theta$) between the two pulses.
For each of the phases the accumulated fluxes are displayed for four time delays
1) $\Delta t$$=$4.84 fs (dark blue),
2) $\Delta t$$=$9.68 fs (light blue),
3) $\Delta t$$=$14.52 fs (yellow) and
4) $\Delta t$$=$ 19.35 fs (red).
The accumulated flux in each vibrational state for a one pulse experiment is shown in
Figure \ref{fig:vibone}.
}
\end{figure*}

In Figure \ref{fig:vibone} the accumulated flux in each vibrational state of desorbing
hydrogen molecule is shown for a one pulse experiment. 
The hydrogen molecule leaves the surface predominantly in the fifth lowest
vibrational excited state ($\nu$$=$4). This observation agrees with the node structure of the
leaving wave function on the product channel (See Figure \ref{fig:wavepacket_product}).
Jackson et al\cite{Persson92} also observed a
vibrational distribution of desorbing hydrogen molecules centered around 
$\nu$$=$4 for an Eley-Rideal reaction of two
hydrogen atoms on a Cu(111)-surface in a collinear quantum mechanical calculation.

The extent of control in the accumulated vibrational flux between the two and one pulse scenario 
is shown in Figure \ref{fig:hhvib1}.
For all the calculations the vibrational distribution is centered around $\nu$$=$4.
An enhancement of flux in the  lowest vibrational states of the hydrogen molecule
($\nu$$\le$3) is correlated with an increases of the total 2PACC yield.
On the other hand an enhancement of the higher lying vibrational states 
($\nu$$\ge$4) is anticorrelated with the  total 2PACC yield. 

\subsection{Li+H/{\bf Cu}(111)}
\begin{figure}[tb!]
\includegraphics[scale=0.5]{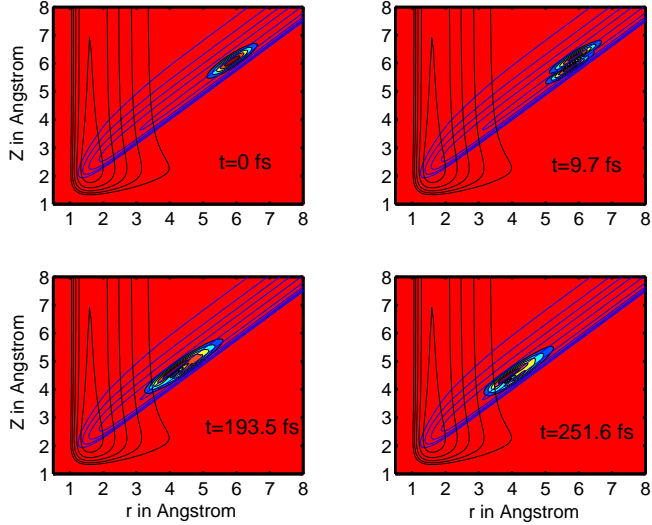}
\caption{\label{fig:reactantlih}
Evolution of the wave packet on the reactant surface for the lithium atom laser impinging 
on a {\bf H}/{\bf Cu}-surface.
The time delay and the relative phase between the two atom laser pulses are 
$\Delta t$$=$9.7fs and $\theta$$=$-$\frac{1}{2}\pi$.
The contours given for -3, -2.5, -2, -1.5, -1 and -0.5 eV for the potential energy surfaces of the product and the reactant channels
are shown with blue and black lines, respectively.}
\end{figure}

Next, the lithium with an adsorb hydrogen Eley-Rideal reaction 
on a {\bf Cu}(111)-surface is studied.
A smaller probability of desorption for  lithium hydride molecule
is expected. This is due to  the reduced energy difference between the product and the reactant  
and a stronger physisorption interaction between the surface and the lithium hydride.
For a single matter-wave pulse, the desorbing flux of lithium hydride is suppressed by $10^{-9}$ 
compared to desorption of hydrogen. 
The total energy of the initial wave packet is -2.4 eV where the initial
kinetic energy of the impinging atom is 0.027 eV.
\begin{figure}[tb!]
\includegraphics[scale=0.5]{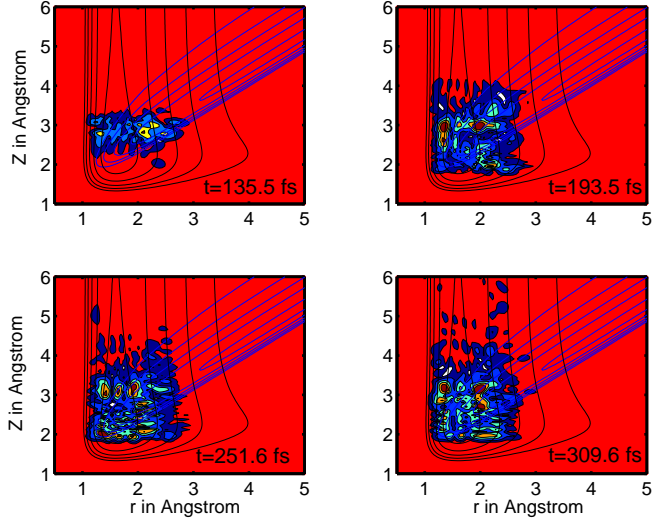}
\caption{\label{fig:productlih}
Evolution of the wave packet on the product PES for lithium atom laser impinging 
on a {\bf H}/{\bf Cu}-surface.
The time delay and the relative phase between the two atom laser pulses are 
$\Delta t$$=$9.7fs and $\theta$$=$-$\frac{1}{2}\pi$.
}
\end{figure}

The evolution of the wave packet on the reactant and the product PES is shown
in Figures \ref{fig:reactantlih} and \ref{fig:productlih}.
The dynamics of the 2PACC spectroscopy with a lithium atom laser is slower 
than the one for a hydrogen 
atom laser since the mass of the impinging atom increases by almost a factor of 7.
As the wave packet on the reactant channel approaches the metal surface,
part of the population is transferred to the product channel. 
The magnitude of population transfer is much smaller for lithium than for hydrogen
as expected. The part of wave packet on the reactant channel which does not react
is primarily trapped on the surface 
due to polarization forces and only a very small part of the wave function scatters back to 
the gas phase. The opposite was observed for the hydrogen case. 

\begin{figure}[tb!]
\includegraphics[scale=0.5]{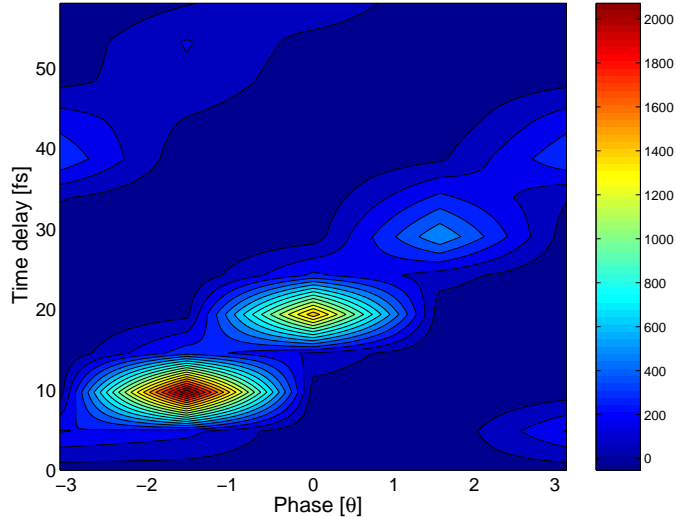}
\caption{\label{fig:2pacclih}The 2PACC signal. The enhancement  
relative to the output of a single pulse
for lithium atom laser applied to a {\bf H}/{\bf Cu}(111)-surface. The control parameters
are  the time delay and phase between the two pulse.
The predicted 2PACC signal is calculated from
the integrated flux along a divided line on the
product channel at $t=484$ fs.
Areas of enhancement are in red and depletion are in blue.
}
\end{figure}

The predicted 2PACC signal is shown in Figure \ref{fig:2pacclih} as a function 
of time delay and phase between the two pulses.
For certain values of time delay and phase between the two pulses
the second pulse induces a large constructive interferences. 
These interferences enhance the probability of desorption for {\bf LiH}
by 2100$\%$ relative to a single-pulse experiment. 

\begin{figure*}[tb!]
\includegraphics[scale=0.9]{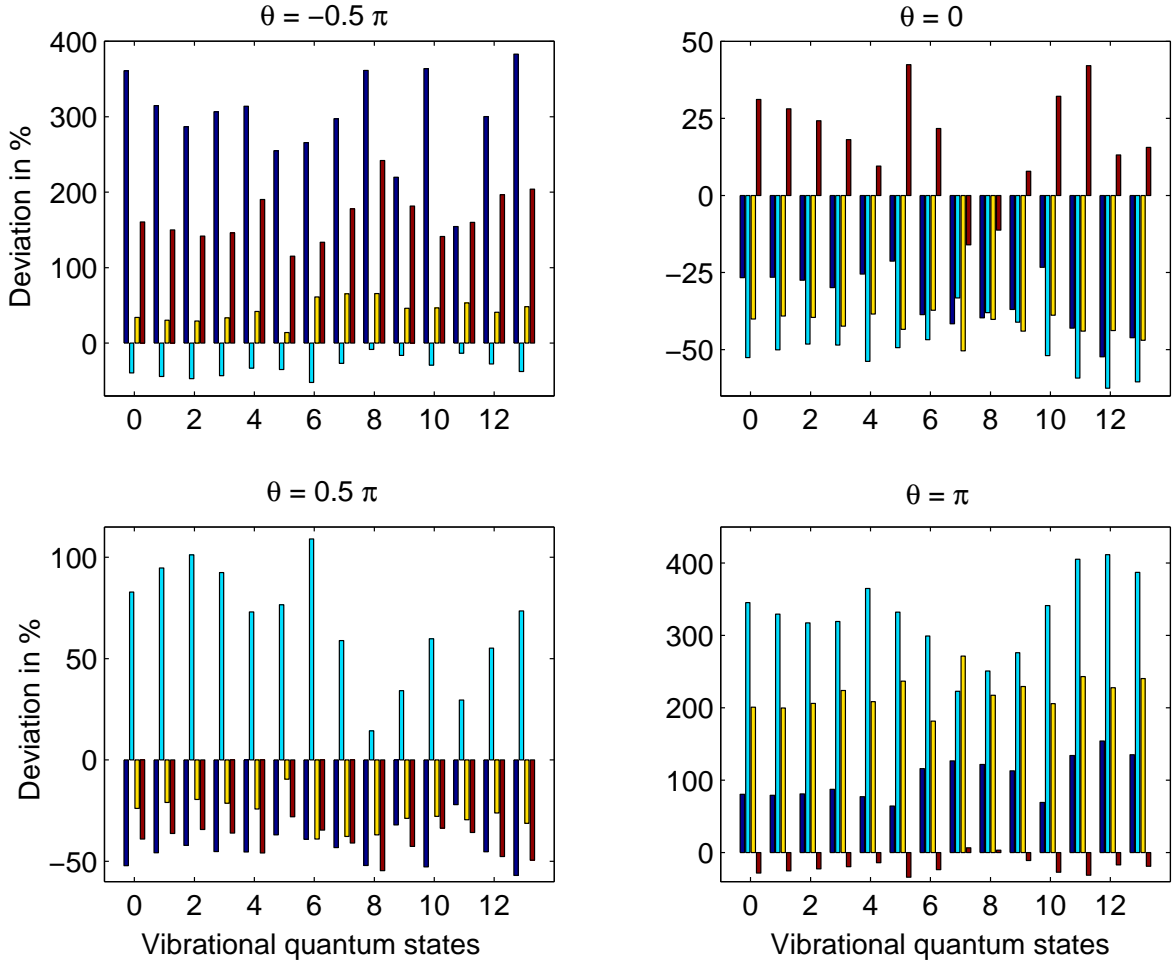}
\caption{\label{fig:lihvib1} 
The percentage deviation in the accumulated flux in the vibrational states 
relative to the single pulse simulation. Four different 
relative phases ($\theta$) between the two pulses are shown. 
For each of the phases the accumulated fluxes are displayed for four time delays 
1) $\Delta t$$=$38.7 fs (dark blue), 
2) $\Delta t$$=$43.5 fs (light blue), 
3) $\Delta t$$=$48.5 fs (yellow) and 
4) $\Delta t$$=$53.2 fs (red).
The accumulated flux in each vibrational state for the single pulse simulation 
is shown in Figure \ref{fig:vibone}.
}
\end{figure*}

The desorbing molecule, {\bf LiH}, is vibrationally excited. 
This is because the bond length {\bf LiH} is considerably longer 
than chemisorption bond between the hydrogen and the {\bf Cu}(111)-surface.
Along the dividing line where the desorbing flux is collected, the fourteen 
lowest vibrational eigenstates of the lithium hydride have been evaluated.
For a single pulse experiment (See Figure \ref{fig:vibone})
a broad distribution of the vibrational states with a maximum
at $\nu$$=$8 is observed. 
In Figure~\ref{fig:lihvib1} the enhancement/suppression in the 
integrated vibrational flux relative to the single-pulse scenario is shown.
The deviation in the integrated flux as a function of the phase reflects
the 2PACC signal. If there is a large enhancement of the 2PACC signal
the accumulated flux in each vibrational state increases.

\section{Summary and Conclusions}
\label{sec:conclusion}

A two-pulse atomic coherent control (2PACC) spectroscopy has been presented in this study.
The coherent properties of an atom laser have been used to control a surface mediated
chemical reaction, $A+B \rar C$,
where the $A$ wave function is generated from an atom laser 
and the target atom $B$ is immobilized by the surface.
Two atom laser pulses are applied creating an interference pattern with each other.
These interferences are the essence of the coherent control of the chemical reaction.
The two pulses are necessary since the initial relative phase between the atom $A$
and atom $B$ on the surface is arbitrary\cite{PBrumer92,PBrumer86}.
The control knobs used were the time delay and a relative phase between two 
atom laser pulses. A significant enhancement of the 2PACC signal
relative to single atom laser pulse was obtained.

The Eley-Rideal reaction forming {\bf H}$_2$ is highly exothermic whereas 
the recombination reaction leading to {\bf LiH} is  almost thermoneutral.  
Using a coherent matter-wave source enables us to coherently control the recombination
reaction.
The 2PACC spectroscopy opens up reaction channels which are almost closed to
incoherent sources.
The reaction yield by Eley-Rideal reaction of lithium atom laser with adsorbed hydrogen
atom has been enhanced by more than 2100$\%$ compared to a one-pulse reference. 
Here, we manipulated the wave function of one atom transferred from a trapped state
in the BEC to an untrapped and propagating state. 
The coherent properties of interest are projected onto this wave function.
In this study a double Gaussians wave function was used where the source of coherent
control is the time delay and relative phase between the two Gaussian wave functions.
The design of the output coupler is the experimental "bottleneck" in coherent control
with matter-waves. In analogy with the optical pulse shaper\cite{AMWeiner88,KANelson91}, a more elaborated matter-wave pulse shaper controlling the amplitude 
and phase of the matter-wave would eventually lead 
to a reaction probability of one.  

The design of the wave function could further be used to direct the leaving 
molecule into a specific rovibrational state.
The 2PACC spectroscopy could also be employed for endothermic reaction such as Eley-Rideal 
reaction between a adsorbed hydrogen atom and sodium (rubidium, cesium). 

Additional coherent manipulations could be possible if
the atomic matter-wave pulse is  replaced
by a molecular pulse where the source is a molecular Bose-Einstein condensate.
For a heterogeneous molecular matter-wave the branching ratio between 
the two reaction channels could be controlled. 
From the prospective of coherent control methodology
the current study demonstrates control of a recombination reaction. 
This is in contrast to optical coherent control which has been almost exclusively limited 
to the control of unimolecular reactions such as photo-dissociation.
 
Control of a recombination reaction is another direction
where both atoms come from the same BEC source. 
In this case their initial relative phase is well defined while the surface
serves to break the symmetry. 
Since both sources are the coherent the reaction product is also  coherent.
The ultimate goal is to design the wave functions of the atom 
lasers such that the reaction probability is unity for one specific 
rovibrational states of the product.
This would lead to the formation of a stable molecular BEC.
Control on these lines is under investigation.

\begin{acknowledgments}

We would like to thank Christiane Koch for proofreading the manuscript.
SJ thanks Marie Curie Fellowship Organization.
This work was supported by the Israel Science Foundation
and the European Research and Training Network
{\em Cold Molecules: Formation, Trapping and Dynamics}.
The Fritz Haber Center is supported by the Minerva 
Gesellschaft f\"ur die Forschung, GmbH M\"unchen, Germany.

\end{acknowledgments}

\appendix

\section{\label{app:a}Vibrational Eigenstates}

Vibrational eigenstates, $\chi_n$,  are solutions of the one-dimensional, 
one channel time-independent Schr\"{o}dinger equation  for a fixed distance,
$Z_{flux}$, from the surface 
\beq
{\bm H}_{PP} (r;Z_{flux}) \chi_n (r;Z_{flux}) 
       = E_n(Z_{flux}) \chi_n (r;Z_{flux}) \;\;.
\eeq
Asymptotically when $Z_{flux}\rar\infty$, the eigenstates represent the vibrational 
states of the diatomic molecule in the gas phase.
The method for evaluating the eigenfunction is based on propagating 
a trial wave function according to the time dependent Schr\"{o}dinger equation 
in imaginary time $\tau=i t$.
A Gaussian filter\cite{ADHammerich89} has been used to project 
the eigenstates out in an energy region around an initial guess, $\epsilon$,
\beq
\chi_{trial}(r,\tau)=
     e^{-4 ({\bm H}-\epsilon)^2 \tau/\hbar\Delta E} \chi_{trial}(r,0)\;\;.
\label{eq:gaussfilter}
\eeq
The parameter $\tau$ has the dimension of time while $\Delta E$ is
the energy range covered by the numerical procedure which appears as a 
scaling factor for the normalization of the Hamiltonian in the Newtonian
time propagation.
This procedure can be interpreted as a relaxation of the ground stated of a
modified Hamiltonian ${\bm H}^*=4({\bm H}-\epsilon)^2/\Delta E$.
Convergence onto a specific target eigenstate can be estimated from 
the energy dispersion relation
\beq
D(\tau)=\sqrt{ \bra \chi(\tau)\mid {\bm H}^2 \mid \chi(\tau) \ket
             - \bra \chi(\tau)\mid {\bm H} \mid \chi(\tau) \ket^2 }
\eeq
which decreases uniformly by  increasing the "purity" of an eigenfunction
and vanishes in the limit of an exact eigenstate.
If the energy dispersion, $D(\tau)$ is less than 1.0E-6
the propagation is stopped and the wave function is a 
vibrational eigenstate of the diatomic molecule, $\chi_n$ for a fixed
distance to the surface.


\begin{thebibliography}{62}
\expandafter\ifx\csname natexlab\endcsname\relax\def\natexlab#1{#1}\fi
\expandafter\ifx\csname bibnamefont\endcsname\relax
  \def\bibnamefont#1{#1}\fi
\expandafter\ifx\csname bibfnamefont\endcsname\relax
  \def\bibfnamefont#1{#1}\fi
\expandafter\ifx\csname citenamefont\endcsname\relax
  \def\citenamefont#1{#1}\fi
\expandafter\ifx\csname url\endcsname\relax
  \def\url#1{\texttt{#1}}\fi
\expandafter\ifx\csname urlprefix\endcsname\relax\def\urlprefix{URL }\fi
\providecommand{\bibinfo}[2]{#2}
\providecommand{\eprint}[2][]{\url{#2}}

\bibitem[{\citenamefont{Rice}(2001)}]{SRice01}
\bibinfo{author}{\bibfnamefont{S.~A.} \bibnamefont{Rice}},
  \bibinfo{journal}{Nature} \textbf{\bibinfo{volume}{409}},
  \bibinfo{pages}{422} (\bibinfo{year}{2001}).

\bibitem[{\citenamefont{Gordon and Rice}(1997)}]{RJGordon97}
\bibinfo{author}{\bibfnamefont{R.~J.} \bibnamefont{Gordon}} \bibnamefont{and}
  \bibinfo{author}{\bibfnamefont{S.~A.} \bibnamefont{Rice}},
  \bibinfo{journal}{Annu. Rev. Phys. Chem.} \textbf{\bibinfo{volume}{48}},
  \bibinfo{pages}{601} (\bibinfo{year}{1997}).

\bibitem[{\citenamefont{Anderson et~al.}(1995)\citenamefont{Anderson, Matthews,
  Wieman, and Cornell}}]{MHAnderson95}
\bibinfo{author}{\bibfnamefont{M.~H.} \bibnamefont{Anderson}},
  \bibinfo{author}{\bibfnamefont{J.~R.} \bibnamefont{Matthews}},
  \bibinfo{author}{\bibfnamefont{C.~E.} \bibnamefont{Wieman}},
  \bibnamefont{and} \bibinfo{author}{\bibfnamefont{E.~A.}
  \bibnamefont{Cornell}}, \bibinfo{journal}{Science}
  \textbf{\bibinfo{volume}{269}}, \bibinfo{pages}{198} (\bibinfo{year}{1995}).

\bibitem[{\citenamefont{Davis et~al.}(1995)\citenamefont{Davis, Mewes, Andrews,
  van Druten, Durfee, Kurn, and Ketterle}}]{KBDavis95}
\bibinfo{author}{\bibfnamefont{K.~B.} \bibnamefont{Davis}},
  \bibinfo{author}{\bibfnamefont{M.-O.} \bibnamefont{Mewes}},
  \bibinfo{author}{\bibfnamefont{M.~R.} \bibnamefont{Andrews}},
  \bibinfo{author}{\bibfnamefont{N.~J.} \bibnamefont{van Druten}},
  \bibinfo{author}{\bibfnamefont{D.~S.} \bibnamefont{Durfee}},
  \bibinfo{author}{\bibfnamefont{D.~M.} \bibnamefont{Kurn}}, \bibnamefont{and}
  \bibinfo{author}{\bibfnamefont{W.}~\bibnamefont{Ketterle}},
  \bibinfo{journal}{Phys. Rev. Lett.} \textbf{\bibinfo{volume}{75}},
  \bibinfo{pages}{3969} (\bibinfo{year}{1995}).

\bibitem[{\citenamefont{Bradley et~al.}(1995)\citenamefont{Bradley, Sackett,
  Tollett, and Hulet}}]{CCBradley95}
\bibinfo{author}{\bibfnamefont{C.~C.} \bibnamefont{Bradley}},
  \bibinfo{author}{\bibfnamefont{C.~A.} \bibnamefont{Sackett}},
  \bibinfo{author}{\bibfnamefont{J.~J.} \bibnamefont{Tollett}},
  \bibnamefont{and} \bibinfo{author}{\bibfnamefont{R.~G.} \bibnamefont{Hulet}},
  \bibinfo{journal}{Phys. Rev. Lett.} \textbf{\bibinfo{volume}{75}},
  \bibinfo{pages}{1687} (\bibinfo{year}{1995}).

\bibitem[{\citenamefont{Bradley et~al.}(1997)\citenamefont{Bradley, Sackett,
  and Hulet}}]{CCBradley97}
\bibinfo{author}{\bibfnamefont{C.~C.} \bibnamefont{Bradley}},
  \bibinfo{author}{\bibfnamefont{C.~A.} \bibnamefont{Sackett}},
  \bibnamefont{and} \bibinfo{author}{\bibfnamefont{R.~G.} \bibnamefont{Hulet}},
  \bibinfo{journal}{Phys. Rev. Lett.} \textbf{\bibinfo{volume}{78}},
  \bibinfo{pages}{985} (\bibinfo{year}{1997}).

\bibitem[{\citenamefont{Mewes et~al.}(1997)\citenamefont{Mewes, Andrews, Kurn,
  Durfee, Townsend, and Ketterle}}]{MOMewes97}
\bibinfo{author}{\bibfnamefont{M.-O.} \bibnamefont{Mewes}},
  \bibinfo{author}{\bibfnamefont{M.~R.} \bibnamefont{Andrews}},
  \bibinfo{author}{\bibfnamefont{D.~M.} \bibnamefont{Kurn}},
  \bibinfo{author}{\bibfnamefont{D.~S.} \bibnamefont{Durfee}},
  \bibinfo{author}{\bibfnamefont{C.~G.} \bibnamefont{Townsend}},
  \bibnamefont{and} \bibinfo{author}{\bibfnamefont{W.}~\bibnamefont{Ketterle}},
  \bibinfo{journal}{Phys. Rev. Lett.} \textbf{\bibinfo{volume}{78}},
  \bibinfo{pages}{582} (\bibinfo{year}{1997}).

\bibitem[{\citenamefont{Anderson and Kasevich}(1998)}]{BPAnderson98}
\bibinfo{author}{\bibfnamefont{B.~P.} \bibnamefont{Anderson}} \bibnamefont{and}
  \bibinfo{author}{\bibfnamefont{M.~A.} \bibnamefont{Kasevich}},
  \bibinfo{journal}{Science} \textbf{\bibinfo{volume}{282}},
  \bibinfo{pages}{1686} (\bibinfo{year}{1998}).

\bibitem[{\citenamefont{Hagley et~al.}(1999)\citenamefont{Hagley, Deng, Kozuma,
  Wen, Helmerson, Rolston, and Phillips}}]{EWHagley99}
\bibinfo{author}{\bibfnamefont{E.~W.} \bibnamefont{Hagley}},
  \bibinfo{author}{\bibfnamefont{L.}~\bibnamefont{Deng}},
  \bibinfo{author}{\bibfnamefont{M.}~\bibnamefont{Kozuma}},
  \bibinfo{author}{\bibfnamefont{J.}~\bibnamefont{Wen}},
  \bibinfo{author}{\bibfnamefont{K.}~\bibnamefont{Helmerson}},
  \bibinfo{author}{\bibfnamefont{S.~L.} \bibnamefont{Rolston}},
  \bibnamefont{and} \bibinfo{author}{\bibfnamefont{W.~D.}
  \bibnamefont{Phillips}}, \bibinfo{journal}{Science}
  \textbf{\bibinfo{volume}{283}}, \bibinfo{pages}{1706} (\bibinfo{year}{1999}).

\bibitem[{\citenamefont{Bloch et~al.}(1999)\citenamefont{Bloch, H{\"{a}}nsch,
  and Esslinger}}]{IBloch99}
\bibinfo{author}{\bibfnamefont{I.}~\bibnamefont{Bloch}},
  \bibinfo{author}{\bibfnamefont{T.~W.} \bibnamefont{H{\"{a}}nsch}},
  \bibnamefont{and}
  \bibinfo{author}{\bibfnamefont{T.}~\bibnamefont{Esslinger}},
  \bibinfo{journal}{Phys. Rev. Lett.} \textbf{\bibinfo{volume}{82}},
  \bibinfo{pages}{3008} (\bibinfo{year}{1999}).

\bibitem[{\citenamefont{Chikkatur et~al.}(2002)\citenamefont{Chikkatur, Shin,
  Leanhardt, Kielpinski, Tsikata, Gustavson, Pritchard, and
  Ketterle}}]{APChikkatur02}
\bibinfo{author}{\bibfnamefont{A.~P.} \bibnamefont{Chikkatur}},
  \bibinfo{author}{\bibfnamefont{Y.}~\bibnamefont{Shin}},
  \bibinfo{author}{\bibfnamefont{A.~E.} \bibnamefont{Leanhardt}},
  \bibinfo{author}{\bibfnamefont{D.}~\bibnamefont{Kielpinski}},
  \bibinfo{author}{\bibfnamefont{E.}~\bibnamefont{Tsikata}},
  \bibinfo{author}{\bibfnamefont{T.~L.} \bibnamefont{Gustavson}},
  \bibinfo{author}{\bibfnamefont{D.~E.} \bibnamefont{Pritchard}},
  \bibnamefont{and} \bibinfo{author}{\bibfnamefont{W.}~\bibnamefont{Ketterle}},
  \bibinfo{journal}{Nature} \textbf{\bibinfo{volume}{296}},
  \bibinfo{pages}{2193} (\bibinfo{year}{2002}).

\bibitem[{\citenamefont{K{\"{o}}hl et~al.}(2002)\citenamefont{K{\"{o}}hl,
  H{\"{a}}nsch, and Esslinger}}]{MKohl02}
\bibinfo{author}{\bibfnamefont{M.}~\bibnamefont{K{\"{o}}hl}},
  \bibinfo{author}{\bibfnamefont{T.~W.} \bibnamefont{H{\"{a}}nsch}},
  \bibnamefont{and}
  \bibinfo{author}{\bibfnamefont{T.}~\bibnamefont{Esslinger}},
  \bibinfo{journal}{Phys. Rev. A} \textbf{\bibinfo{volume}{65}},
  \bibinfo{pages}{021606} (\bibinfo{year}{2002}).

\bibitem[{\citenamefont{Coq et~al.}(2001)\citenamefont{Coq, Thywissen,
  Rangwala, Gerbier, Richard, Delannoy, Bouyer, and Aspect}}]{YLCoq01}
\bibinfo{author}{\bibfnamefont{Y.~L.} \bibnamefont{Coq}},
  \bibinfo{author}{\bibfnamefont{J.~H.} \bibnamefont{Thywissen}},
  \bibinfo{author}{\bibfnamefont{S.~A.} \bibnamefont{Rangwala}},
  \bibinfo{author}{\bibfnamefont{F.}~\bibnamefont{Gerbier}},
  \bibinfo{author}{\bibfnamefont{S.}~\bibnamefont{Richard}},
  \bibinfo{author}{\bibfnamefont{G.}~\bibnamefont{Delannoy}},
  \bibinfo{author}{\bibfnamefont{P.}~\bibnamefont{Bouyer}}, \bibnamefont{and}
  \bibinfo{author}{\bibfnamefont{A.}~\bibnamefont{Aspect}},
  \bibinfo{journal}{Phys. Rev. Lett.} \textbf{\bibinfo{volume}{87}},
  \bibinfo{pages}{170403} (\bibinfo{year}{2001}).

\bibitem[{\citenamefont{J{\o}rgensen and Kosloff}(2003)}]{2pacc:1}
\bibinfo{author}{\bibfnamefont{S.}~\bibnamefont{J{\o}rgensen}}
  \bibnamefont{and} \bibinfo{author}{\bibfnamefont{R.}~\bibnamefont{Kosloff}},
  \bibinfo{journal}{Accepted for publication in Surf. Sci.}
  (\bibinfo{year}{2003}).

\bibitem[{\citenamefont{J{\o}rgensen and Kosloff}(2002)}]{2pacc:3}
\bibinfo{author}{\bibfnamefont{S.}~\bibnamefont{J{\o}rgensen}}
  \bibnamefont{and} \bibinfo{author}{\bibfnamefont{R.}~\bibnamefont{Kosloff}},
  in \emph{\bibinfo{booktitle}{Interactions in Ultracold Gases: From Atoms to
  Molecules.}}, edited by
  \bibinfo{editor}{\bibfnamefont{M.}~\bibnamefont{Weidem{\"u}ller}}
  \bibnamefont{and}
  \bibinfo{editor}{\bibfnamefont{C.}~\bibnamefont{Zimmermann}}
  (\bibinfo{publisher}{Wiley}, \bibinfo{address}{Berlin},
  \bibinfo{year}{2002}).

\bibitem[{\citenamefont{Weber et~al.}(2003)\citenamefont{Weber, Herbig, Mark,
  N{\"{a}}gerl, , and Grimm}}]{RGrimm02}
\bibinfo{author}{\bibfnamefont{T.}~\bibnamefont{Weber}},
  \bibinfo{author}{\bibfnamefont{J.}~\bibnamefont{Herbig}},
  \bibinfo{author}{\bibfnamefont{M.}~\bibnamefont{Mark}},
  \bibinfo{author}{\bibfnamefont{H.-C.} \bibnamefont{N{\"{a}}gerl}}, ,
  \bibnamefont{and} \bibinfo{author}{\bibfnamefont{R.}~\bibnamefont{Grimm}},
  \bibinfo{journal}{Science} \textbf{\bibinfo{volume}{299}},
  \bibinfo{pages}{232} (\bibinfo{year}{2003}).

\bibitem[{\citenamefont{Cornish et~al.}(2000)\citenamefont{Cornish, Claussen,
  Roberts, Cornell, and Wieman}}]{SLCornish00}
\bibinfo{author}{\bibfnamefont{S.~L.} \bibnamefont{Cornish}},
  \bibinfo{author}{\bibfnamefont{N.~R.} \bibnamefont{Claussen}},
  \bibinfo{author}{\bibfnamefont{J.~L.} \bibnamefont{Roberts}},
  \bibinfo{author}{\bibfnamefont{E.~A.} \bibnamefont{Cornell}},
  \bibnamefont{and} \bibinfo{author}{\bibfnamefont{C.~E.}
  \bibnamefont{Wieman}}, \bibinfo{journal}{Phys. Rev. Lett.}
  \textbf{\bibinfo{volume}{85}}, \bibinfo{pages}{1795} (\bibinfo{year}{2000}).

\bibitem[{\citenamefont{Schreck et~al.}(2001)\citenamefont{Schreck, Khaykovich,
  Corwin, Ferrari, Bourdel, Cubizolles, and Salomon}}]{FSchreck01}
\bibinfo{author}{\bibfnamefont{F.}~\bibnamefont{Schreck}},
  \bibinfo{author}{\bibfnamefont{L.}~\bibnamefont{Khaykovich}},
  \bibinfo{author}{\bibfnamefont{K.~L.} \bibnamefont{Corwin}},
  \bibinfo{author}{\bibfnamefont{G.}~\bibnamefont{Ferrari}},
  \bibinfo{author}{\bibfnamefont{T.}~\bibnamefont{Bourdel}},
  \bibinfo{author}{\bibfnamefont{J.}~\bibnamefont{Cubizolles}},
  \bibnamefont{and} \bibinfo{author}{\bibfnamefont{C.}~\bibnamefont{Salomon}},
  \bibinfo{journal}{Phys. Rev. Lett.} \textbf{\bibinfo{volume}{87}},
  \bibinfo{pages}{080403} (\bibinfo{year}{2001}).

\bibitem[{\citenamefont{Modugnu et~al.}(2001)\citenamefont{Modugnu, Ferrari,
  Roati, Brecha, Simoni, and Inguscio}}]{Modugno01}
\bibinfo{author}{\bibfnamefont{G.}~\bibnamefont{Modugnu}},
  \bibinfo{author}{\bibfnamefont{G.}~\bibnamefont{Ferrari}},
  \bibinfo{author}{\bibfnamefont{G.}~\bibnamefont{Roati}},
  \bibinfo{author}{\bibfnamefont{R.~J.} \bibnamefont{Brecha}},
  \bibinfo{author}{\bibfnamefont{A.}~\bibnamefont{Simoni}}, \bibnamefont{and}
  \bibinfo{author}{\bibfnamefont{M.}~\bibnamefont{Inguscio}},
  \bibinfo{journal}{Science} \textbf{\bibinfo{volume}{294}},
  \bibinfo{pages}{1320} (\bibinfo{year}{2001}).

\bibitem[{\citenamefont{Fried et~al.}(1998)\citenamefont{Fried, Killian,
  Willmann, Landhuis, Moss, Kleppner, and Greytak}}]{DGFried98}
\bibinfo{author}{\bibfnamefont{D.~G.} \bibnamefont{Fried}},
  \bibinfo{author}{\bibfnamefont{T.~C.} \bibnamefont{Killian}},
  \bibinfo{author}{\bibfnamefont{L.}~\bibnamefont{Willmann}},
  \bibinfo{author}{\bibfnamefont{D.}~\bibnamefont{Landhuis}},
  \bibinfo{author}{\bibfnamefont{S.~C.} \bibnamefont{Moss}},
  \bibinfo{author}{\bibfnamefont{D.}~\bibnamefont{Kleppner}}, \bibnamefont{and}
  \bibinfo{author}{\bibfnamefont{T.~J.} \bibnamefont{Greytak}},
  \bibinfo{journal}{Phys. Rev. Lett.} \textbf{\bibinfo{volume}{81}},
  \bibinfo{pages}{3811} (\bibinfo{year}{1998}).

\bibitem[{\citenamefont{Santos et~al.}(2001)\citenamefont{Santos, L\'eonard,
  Wang, Barrelet, Perales, Rasel, Unnikrishnan, Leduc, and
  Cohen-Tannoudji}}]{FPDSantos01}
\bibinfo{author}{\bibfnamefont{F.~P.~D.} \bibnamefont{Santos}},
  \bibinfo{author}{\bibfnamefont{J.}~\bibnamefont{L\'eonard}},
  \bibinfo{author}{\bibfnamefont{J.}~\bibnamefont{Wang}},
  \bibinfo{author}{\bibfnamefont{C.~J.} \bibnamefont{Barrelet}},
  \bibinfo{author}{\bibfnamefont{F.}~\bibnamefont{Perales}},
  \bibinfo{author}{\bibfnamefont{E.}~\bibnamefont{Rasel}},
  \bibinfo{author}{\bibfnamefont{C.~S.} \bibnamefont{Unnikrishnan}},
  \bibinfo{author}{\bibfnamefont{M.}~\bibnamefont{Leduc}}, \bibnamefont{and}
  \bibinfo{author}{\bibfnamefont{C.}~\bibnamefont{Cohen-Tannoudji}},
  \bibinfo{journal}{Phys. Rev. Lett.} \textbf{\bibinfo{volume}{86}},
  \bibinfo{pages}{3459} (\bibinfo{year}{2001}).

\bibitem[{\citenamefont{Robert et~al.}(2001)\citenamefont{Robert, Sirjean,
  Browaeys, Poupard, Nowak, Boiron, Westbrook, and Aspect}}]{ARobert01}
\bibinfo{author}{\bibfnamefont{A.}~\bibnamefont{Robert}},
  \bibinfo{author}{\bibfnamefont{O.}~\bibnamefont{Sirjean}},
  \bibinfo{author}{\bibfnamefont{A.}~\bibnamefont{Browaeys}},
  \bibinfo{author}{\bibfnamefont{J.}~\bibnamefont{Poupard}},
  \bibinfo{author}{\bibfnamefont{S.}~\bibnamefont{Nowak}},
  \bibinfo{author}{\bibfnamefont{D.}~\bibnamefont{Boiron}},
  \bibinfo{author}{\bibfnamefont{C.~I.} \bibnamefont{Westbrook}},
  \bibnamefont{and} \bibinfo{author}{\bibfnamefont{A.}~\bibnamefont{Aspect}},
  \bibinfo{journal}{Science} \textbf{\bibinfo{volume}{292}},
  \bibinfo{pages}{461} (\bibinfo{year}{2001}).

\bibitem[{\citenamefont{Scherer et~al.}(1990)\citenamefont{Scherer, Ruggiero,
  Du, and Fleming}}]{NFScherer90}
\bibinfo{author}{\bibfnamefont{N.~F.} \bibnamefont{Scherer}},
  \bibinfo{author}{\bibfnamefont{A.~J.} \bibnamefont{Ruggiero}},
  \bibinfo{author}{\bibfnamefont{M.}~\bibnamefont{Du}}, \bibnamefont{and}
  \bibinfo{author}{\bibfnamefont{G.~R.} \bibnamefont{Fleming}},
  \bibinfo{journal}{J. Chem. Phys.} \textbf{\bibinfo{volume}{93}},
  \bibinfo{pages}{856} (\bibinfo{year}{1990}).

\bibitem[{\citenamefont{Scherer et~al.}(1991)\citenamefont{Scherer, Carlson,
  Matro, Du, Ruggiero, Romero-Rochin, Cina, and Fleming}}]{NFScherer91}
\bibinfo{author}{\bibfnamefont{N.~F.} \bibnamefont{Scherer}},
  \bibinfo{author}{\bibfnamefont{R.~J.} \bibnamefont{Carlson}},
  \bibinfo{author}{\bibfnamefont{A.}~\bibnamefont{Matro}},
  \bibinfo{author}{\bibfnamefont{M.}~\bibnamefont{Du}},
  \bibinfo{author}{\bibfnamefont{A.~J.} \bibnamefont{Ruggiero}},
  \bibinfo{author}{\bibfnamefont{V.}~\bibnamefont{Romero-Rochin}},
  \bibinfo{author}{\bibfnamefont{J.~A.} \bibnamefont{Cina}}, \bibnamefont{and}
  \bibinfo{author}{\bibfnamefont{G.~R.} \bibnamefont{Fleming}},
  \bibinfo{journal}{J. Chem. Phys.} \textbf{\bibinfo{volume}{95}},
  \bibinfo{pages}{1487} (\bibinfo{year}{1991}).

\bibitem[{\citenamefont{Kosloff et~al.}(1992)\citenamefont{Kosloff, Hammerich,
  and Tannor}}]{RKosloff92}
\bibinfo{author}{\bibfnamefont{R.}~\bibnamefont{Kosloff}},
  \bibinfo{author}{\bibfnamefont{A.~D.} \bibnamefont{Hammerich}},
  \bibnamefont{and} \bibinfo{author}{\bibfnamefont{D.}~\bibnamefont{Tannor}},
  \bibinfo{journal}{Phys. Rev. Lett.} \textbf{\bibinfo{volume}{69}},
  \bibinfo{pages}{2171} (\bibinfo{year}{1992}).

\bibitem[{\citenamefont{Bartana et~al.}(1993)\citenamefont{Bartana, Kosloff,
  and Tannor}}]{ABartana93}
\bibinfo{author}{\bibfnamefont{A.}~\bibnamefont{Bartana}},
  \bibinfo{author}{\bibfnamefont{R.}~\bibnamefont{Kosloff}}, \bibnamefont{and}
  \bibinfo{author}{\bibfnamefont{D.}~\bibnamefont{Tannor}},
  \bibinfo{journal}{J. Chem. Phys.} \textbf{\bibinfo{volume}{99}},
  \bibinfo{pages}{196} (\bibinfo{year}{1993}).

\bibitem[{\citenamefont{Abrashkevich et~al.}(1998)\citenamefont{Abrashkevich,
  Shapiro, and Brumer}}]{PBrumer98}
\bibinfo{author}{\bibfnamefont{A.}~\bibnamefont{Abrashkevich}},
  \bibinfo{author}{\bibfnamefont{M.}~\bibnamefont{Shapiro}}, \bibnamefont{and}
  \bibinfo{author}{\bibfnamefont{P.}~\bibnamefont{Brumer}},
  \bibinfo{journal}{Phys. Rev. Lett.} \textbf{\bibinfo{volume}{81}},
  \bibinfo{pages}{3789} (\bibinfo{year}{1998}).

\bibitem[{\citenamefont{Jones et~al.}(2003)\citenamefont{Jones, Vale, Sahagun,
  Hall, and Hinds}}]{MPAJones03}
\bibinfo{author}{\bibfnamefont{M.~P.~A.} \bibnamefont{Jones}},
  \bibinfo{author}{\bibfnamefont{C.~J.} \bibnamefont{Vale}},
  \bibinfo{author}{\bibfnamefont{D.}~\bibnamefont{Sahagun}},
  \bibinfo{author}{\bibfnamefont{B.~V.} \bibnamefont{Hall}}, \bibnamefont{and}
  \bibinfo{author}{\bibfnamefont{E.~A.} \bibnamefont{Hinds}},
  \bibinfo{journal}{quant-ph/0301018}  (\bibinfo{year}{2003}).

\bibitem[{\citenamefont{Folman et~al.}(2002)\citenamefont{Folman, Kr{\"{u}}ger,
  Henkel, and Schmiedmayer}}]{RFolman02}
\bibinfo{author}{\bibfnamefont{R.}~\bibnamefont{Folman}},
  \bibinfo{author}{\bibfnamefont{P.}~\bibnamefont{Kr{\"{u}}ger}},
  \bibinfo{author}{\bibfnamefont{C.}~\bibnamefont{Henkel}}, \bibnamefont{and}
  \bibinfo{author}{\bibfnamefont{J.}~\bibnamefont{Schmiedmayer}},
  \bibinfo{journal}{Adv. Atom. Mol. Opt. Phys} \textbf{\bibinfo{volume}{48}},
  \bibinfo{pages}{263} (\bibinfo{year}{2002}).

\bibitem[{\citenamefont{H{\"{a}}nsel et~al.}(2001)\citenamefont{H{\"{a}}nsel,
  Hommelhoff, H{\"{a}}nsch, and Reichel}}]{WHansel01}
\bibinfo{author}{\bibfnamefont{W.}~\bibnamefont{H{\"{a}}nsel}},
  \bibinfo{author}{\bibfnamefont{P.}~\bibnamefont{Hommelhoff}},
  \bibinfo{author}{\bibfnamefont{T.~W.} \bibnamefont{H{\"{a}}nsch}},
  \bibnamefont{and} \bibinfo{author}{\bibfnamefont{J.}~\bibnamefont{Reichel}},
  \bibinfo{journal}{Nature} \textbf{\bibinfo{volume}{413}},
  \bibinfo{pages}{498} (\bibinfo{year}{2001}).

\bibitem[{\citenamefont{Ott et~al.}(2001)\citenamefont{Ott, Fort{\'{a}}gh,
  Schlotterbeck, Grossmann, and Zimmerman}}]{HOtt01}
\bibinfo{author}{\bibfnamefont{H.}~\bibnamefont{Ott}},
  \bibinfo{author}{\bibfnamefont{J.}~\bibnamefont{Fort{\'{a}}gh}},
  \bibinfo{author}{\bibfnamefont{G.}~\bibnamefont{Schlotterbeck}},
  \bibinfo{author}{\bibfnamefont{A.}~\bibnamefont{Grossmann}},
  \bibnamefont{and}
  \bibinfo{author}{\bibfnamefont{C.}~\bibnamefont{Zimmerman}},
  \bibinfo{journal}{Phys. Rev. Lett.} \textbf{\bibinfo{volume}{87}},
  \bibinfo{pages}{230401} (\bibinfo{year}{2001}).

\bibitem[{\citenamefont{Leanhardt et~al.}(2002)\citenamefont{Leanhardt,
  Chikkatur, Kielpinski, Shin, Gustavson, Ketterle, and
  Pritchard}}]{AELeanhardt02}
\bibinfo{author}{\bibfnamefont{A.~E.} \bibnamefont{Leanhardt}},
  \bibinfo{author}{\bibfnamefont{A.~P.} \bibnamefont{Chikkatur}},
  \bibinfo{author}{\bibfnamefont{D.}~\bibnamefont{Kielpinski}},
  \bibinfo{author}{\bibfnamefont{Y.}~\bibnamefont{Shin}},
  \bibinfo{author}{\bibfnamefont{T.~L.} \bibnamefont{Gustavson}},
  \bibinfo{author}{\bibfnamefont{W.}~\bibnamefont{Ketterle}}, \bibnamefont{and}
  \bibinfo{author}{\bibfnamefont{D.~E.} \bibnamefont{Pritchard}},
  \bibinfo{journal}{Phys. Rev. Lett.} \textbf{\bibinfo{volume}{89}},
  \bibinfo{pages}{040401} (\bibinfo{year}{2002}).

\bibitem[{\citenamefont{Geisen et~al.}(1985)\citenamefont{Geisen, Hage,
  Himpsel, Reiss, and Steinmann}}]{Giesen85}
\bibinfo{author}{\bibfnamefont{K.}~\bibnamefont{Geisen}},
  \bibinfo{author}{\bibfnamefont{F.}~\bibnamefont{Hage}},
  \bibinfo{author}{\bibfnamefont{F.~J.} \bibnamefont{Himpsel}},
  \bibinfo{author}{\bibfnamefont{H.~J.} \bibnamefont{Reiss}}, \bibnamefont{and}
  \bibinfo{author}{\bibfnamefont{W.}~\bibnamefont{Steinmann}},
  \bibinfo{journal}{Phys. Rev. Lett.} \textbf{\bibinfo{volume}{55}},
  \bibinfo{pages}{300} (\bibinfo{year}{1985}).

\bibitem[{\citenamefont{Fauster and Steinmann}(1995)}]{Fauster95}
\bibinfo{author}{\bibfnamefont{T.}~\bibnamefont{Fauster}} \bibnamefont{and}
  \bibinfo{author}{\bibfnamefont{W.}~\bibnamefont{Steinmann}}, in
  \emph{\bibinfo{booktitle}{Photonic Probes of Surfaces.}}, edited by
  \bibinfo{editor}{\bibfnamefont{P.}~\bibnamefont{Halevi}}
  (\bibinfo{publisher}{North-Holland}, \bibinfo{address}{Amsterdam},
  \bibinfo{year}{1995}), vol.~\bibinfo{volume}{2}, p. \bibinfo{pages}{347}.

\bibitem[{\citenamefont{Harris et~al.}(1997)\citenamefont{Harris, Ge, Lingle,
  McNeill, and Wong}}]{Review97}
\bibinfo{author}{\bibfnamefont{C.~B.} \bibnamefont{Harris}},
  \bibinfo{author}{\bibfnamefont{N.-H.} \bibnamefont{Ge}},
  \bibinfo{author}{\bibfnamefont{R.~L.} \bibnamefont{Lingle}},
  \bibinfo{author}{\bibfnamefont{J.~D.} \bibnamefont{McNeill}},
  \bibnamefont{and} \bibinfo{author}{\bibfnamefont{C.~M.} \bibnamefont{Wong}},
  \bibinfo{journal}{Annu. Rev. Phys. Chem.} \textbf{\bibinfo{volume}{48}},
  \bibinfo{pages}{711} (\bibinfo{year}{1997}).

\bibitem[{\citenamefont{Petek and Ogawa}(2002)}]{Petek02}
\bibinfo{author}{\bibfnamefont{H.}~\bibnamefont{Petek}} \bibnamefont{and}
  \bibinfo{author}{\bibfnamefont{S.}~\bibnamefont{Ogawa}},
  \bibinfo{journal}{Annu. Rev. Phys. Chem.} \textbf{\bibinfo{volume}{53}},
  \bibinfo{pages}{507} (\bibinfo{year}{2002}).

\bibitem[{\citenamefont{Shumay et~al.}(1998)\citenamefont{Shumay, H{\"{o}}fer,
  Reuss, Thomann, Wallauer, and Fauster}}]{shumay98}
\bibinfo{author}{\bibfnamefont{I.~L.} \bibnamefont{Shumay}},
  \bibinfo{author}{\bibfnamefont{U.}~\bibnamefont{H{\"{o}}fer}},
  \bibinfo{author}{\bibfnamefont{C.}~\bibnamefont{Reuss}},
  \bibinfo{author}{\bibfnamefont{U.}~\bibnamefont{Thomann}},
  \bibinfo{author}{\bibfnamefont{W.}~\bibnamefont{Wallauer}}, \bibnamefont{and}
  \bibinfo{author}{\bibfnamefont{T.}~\bibnamefont{Fauster}},
  \bibinfo{journal}{Phys. Rev. B} \textbf{\bibinfo{volume}{58}},
  \bibinfo{pages}{13974} (\bibinfo{year}{1998}).

\bibitem[{\citenamefont{Vondrak et~al.}(2000)\citenamefont{Vondrak, Wang,
  Winget, Cramer, and Zhu}}]{zhu2000}
\bibinfo{author}{\bibfnamefont{T.}~\bibnamefont{Vondrak}},
  \bibinfo{author}{\bibfnamefont{H.}~\bibnamefont{Wang}},
  \bibinfo{author}{\bibfnamefont{P.}~\bibnamefont{Winget}},
  \bibinfo{author}{\bibfnamefont{C.~J.} \bibnamefont{Cramer}},
  \bibnamefont{and} \bibinfo{author}{\bibfnamefont{X.-Y.} \bibnamefont{Zhu}},
  \bibinfo{journal}{J. Am. Chem. Soc.} \textbf{\bibinfo{volume}{122}},
  \bibinfo{pages}{4700} (\bibinfo{year}{2000}).

\bibitem[{\citenamefont{Gahl et~al.}(2000)\citenamefont{Gahl, Ishioka, Zhong,
  Hotzel, and Wolf}}]{Wolf2000}
\bibinfo{author}{\bibfnamefont{C.}~\bibnamefont{Gahl}},
  \bibinfo{author}{\bibfnamefont{K.}~\bibnamefont{Ishioka}},
  \bibinfo{author}{\bibfnamefont{Q.}~\bibnamefont{Zhong}},
  \bibinfo{author}{\bibfnamefont{A.}~\bibnamefont{Hotzel}}, \bibnamefont{and}
  \bibinfo{author}{\bibfnamefont{M.}~\bibnamefont{Wolf}},
  \bibinfo{journal}{Faraday Discuss} \textbf{\bibinfo{volume}{117}},
  \bibinfo{pages}{191} (\bibinfo{year}{2000}).

\bibitem[{\citenamefont{Sha et~al.}(2002)\citenamefont{Sha, Jackson, and
  Lemoine}}]{Lemoine02}
\bibinfo{author}{\bibfnamefont{X.}~\bibnamefont{Sha}},
  \bibinfo{author}{\bibfnamefont{B.}~\bibnamefont{Jackson}}, \bibnamefont{and}
  \bibinfo{author}{\bibfnamefont{D.}~\bibnamefont{Lemoine}},
  \bibinfo{journal}{J. Chem. Phys.} \textbf{\bibinfo{volume}{116}},
  \bibinfo{pages}{7158} (\bibinfo{year}{2002}).

\bibitem[{\citenamefont{Jackson and Lemoine}(2001)}]{Lemoine01}
\bibinfo{author}{\bibfnamefont{B.}~\bibnamefont{Jackson}} \bibnamefont{and}
  \bibinfo{author}{\bibfnamefont{D.}~\bibnamefont{Lemoine}},
  \bibinfo{journal}{J. Chem. Phys.} \textbf{\bibinfo{volume}{114}},
  \bibinfo{pages}{474} (\bibinfo{year}{2001}).

\bibitem[{\citenamefont{Shalashilin et~al.}(1999)\citenamefont{Shalashilin,
  Jackson, and Persson}}]{Persson99}
\bibinfo{author}{\bibfnamefont{D.~V.} \bibnamefont{Shalashilin}},
  \bibinfo{author}{\bibfnamefont{B.}~\bibnamefont{Jackson}}, \bibnamefont{and}
  \bibinfo{author}{\bibfnamefont{M.}~\bibnamefont{Persson}},
  \bibinfo{journal}{J. Chem. Phys.} \textbf{\bibinfo{volume}{110}},
  \bibinfo{pages}{11038} (\bibinfo{year}{1999}).

\bibitem[{\citenamefont{Jackson and Persson}(1992)}]{Persson92}
\bibinfo{author}{\bibfnamefont{B.}~\bibnamefont{Jackson}} \bibnamefont{and}
  \bibinfo{author}{\bibfnamefont{M.}~\bibnamefont{Persson}},
  \bibinfo{journal}{J. Chem. Phys.} \textbf{\bibinfo{volume}{96}},
  \bibinfo{pages}{2378} (\bibinfo{year}{1992}).

\bibitem[{\citenamefont{Ashkenazi et~al.}(1995)\citenamefont{Ashkenazi,
  Kosloff, Ruhman, and Tal-Ezer}}]{Ashkenazi95}
\bibinfo{author}{\bibfnamefont{G.}~\bibnamefont{Ashkenazi}},
  \bibinfo{author}{\bibfnamefont{R.}~\bibnamefont{Kosloff}},
  \bibinfo{author}{\bibfnamefont{S.}~\bibnamefont{Ruhman}}, \bibnamefont{and}
  \bibinfo{author}{\bibfnamefont{H.}~\bibnamefont{Tal-Ezer}},
  \bibinfo{journal}{J. Chem. Phys.} \textbf{\bibinfo{volume}{103}},
  \bibinfo{pages}{10005} (\bibinfo{year}{1995}).

\bibitem[{\citenamefont{Castro et~al.}(1996)\citenamefont{Castro, Drakova,
  Grillo, and Doyen}}]{GDoyen96}
\bibinfo{author}{\bibfnamefont{G.~R.} \bibnamefont{Castro}},
  \bibinfo{author}{\bibfnamefont{D.}~\bibnamefont{Drakova}},
  \bibinfo{author}{\bibfnamefont{M.~E.} \bibnamefont{Grillo}},
  \bibnamefont{and} \bibinfo{author}{\bibfnamefont{G.}~\bibnamefont{Doyen}},
  \bibinfo{journal}{J. Chem. Phys.} \textbf{\bibinfo{volume}{105}},
  \bibinfo{pages}{9640} (\bibinfo{year}{1996}).

\bibitem[{\citenamefont{Darling and Holloway}(1995)}]{GRDarling95}
\bibinfo{author}{\bibfnamefont{G.~R.} \bibnamefont{Darling}} \bibnamefont{and}
  \bibinfo{author}{\bibfnamefont{S.}~\bibnamefont{Holloway}},
  \bibinfo{journal}{Rep. Prog. Phys.} \textbf{\bibinfo{volume}{58}},
  \bibinfo{pages}{1595} (\bibinfo{year}{1995}).

\bibitem[{\citenamefont{Harris et~al.}(1988)\citenamefont{Harris, Holloway,
  Rahman, and Yang}}]{JHarris88}
\bibinfo{author}{\bibfnamefont{J.}~\bibnamefont{Harris}},
  \bibinfo{author}{\bibfnamefont{S.}~\bibnamefont{Holloway}},
  \bibinfo{author}{\bibfnamefont{T.~S.} \bibnamefont{Rahman}},
  \bibnamefont{and} \bibinfo{author}{\bibfnamefont{K.}~\bibnamefont{Yang}},
  \bibinfo{journal}{J. Chem. Phys.} \textbf{\bibinfo{volume}{89}},
  \bibinfo{pages}{4427} (\bibinfo{year}{1988}).

\bibitem[{\citenamefont{Hand and Holloway}(1989)}]{MRHand89}
\bibinfo{author}{\bibfnamefont{M.~R.} \bibnamefont{Hand}} \bibnamefont{and}
  \bibinfo{author}{\bibfnamefont{S.}~\bibnamefont{Holloway}},
  \bibinfo{journal}{J. Chem. Phys.} \textbf{\bibinfo{volume}{91}},
  \bibinfo{pages}{7209} (\bibinfo{year}{1989}).

\bibitem[{\citenamefont{Persson et~al.}(1999)\citenamefont{Persson,
  Str{\"{o}}mquist, Bengtsson, Jackson, Shalashilin, and Hammer}}]{Hammer99}
\bibinfo{author}{\bibfnamefont{M.}~\bibnamefont{Persson}},
  \bibinfo{author}{\bibfnamefont{J.}~\bibnamefont{Str{\"{o}}mquist}},
  \bibinfo{author}{\bibfnamefont{L.}~\bibnamefont{Bengtsson}},
  \bibinfo{author}{\bibfnamefont{B.}~\bibnamefont{Jackson}},
  \bibinfo{author}{\bibfnamefont{D.~V.} \bibnamefont{Shalashilin}},
  \bibnamefont{and} \bibinfo{author}{\bibfnamefont{B.}~\bibnamefont{Hammer}},
  \bibinfo{journal}{J. Chem. Phys.} \textbf{\bibinfo{volume}{110}},
  \bibinfo{pages}{2240} (\bibinfo{year}{1999}).

\bibitem[{\citenamefont{Truong et~al.}(1989)\citenamefont{Truong, Truhlar, and
  Garrett}}]{TNTruong89}
\bibinfo{author}{\bibfnamefont{T.~N.} \bibnamefont{Truong}},
  \bibinfo{author}{\bibfnamefont{D.~G.} \bibnamefont{Truhlar}},
  \bibnamefont{and} \bibinfo{author}{\bibfnamefont{B.~C.}
  \bibnamefont{Garrett}}, \bibinfo{journal}{J. Phys. Chem.}
  \textbf{\bibinfo{volume}{93}}, \bibinfo{pages}{8227} (\bibinfo{year}{1989}).

\bibitem[{\citenamefont{Tully and Billing}(2002)}]{CTully02}
\bibinfo{author}{\bibfnamefont{C.}~\bibnamefont{Tully}} \bibnamefont{and}
  \bibinfo{author}{\bibfnamefont{G.~D.} \bibnamefont{Billing}},
  \bibinfo{journal}{Submitted to Elviser Science}  (\bibinfo{year}{2002}).

\bibitem[{\citenamefont{Kolos and Wolniewicz}(1965)}]{WKolos65}
\bibinfo{author}{\bibfnamefont{W.}~\bibnamefont{Kolos}} \bibnamefont{and}
  \bibinfo{author}{\bibfnamefont{L.}~\bibnamefont{Wolniewicz}},
  \bibinfo{journal}{J. Chem. Phys.} \textbf{\bibinfo{volume}{43}},
  \bibinfo{pages}{2340} (\bibinfo{year}{1965}).

\bibitem[{\citenamefont{Geum et~al.}(2001)\citenamefont{Geum, Jeung,
  Derevianko, Cote, and Dalgarno}}]{NGeum01}
\bibinfo{author}{\bibfnamefont{N.}~\bibnamefont{Geum}},
  \bibinfo{author}{\bibfnamefont{G.-H.} \bibnamefont{Jeung}},
  \bibinfo{author}{\bibfnamefont{A.}~\bibnamefont{Derevianko}},
  \bibinfo{author}{\bibfnamefont{R.}~\bibnamefont{Cote}}, \bibnamefont{and}
  \bibinfo{author}{\bibfnamefont{A.}~\bibnamefont{Dalgarno}},
  \bibinfo{journal}{J. Chem. Phys.} \textbf{\bibinfo{volume}{115}},
  \bibinfo{pages}{5984} (\bibinfo{year}{2001}).

\bibitem[{\citenamefont{Kosloff and Tal-Ezer}(1986)}]{Kosloff86}
\bibinfo{author}{\bibfnamefont{R.}~\bibnamefont{Kosloff}} \bibnamefont{and}
  \bibinfo{author}{\bibfnamefont{H.}~\bibnamefont{Tal-Ezer}},
  \bibinfo{journal}{Chem. Phys. Lett.} \textbf{\bibinfo{volume}{127}},
  \bibinfo{pages}{223} (\bibinfo{year}{1986}).

\bibitem[{\citenamefont{Kosloff}(1994)}]{Kosloff94b}
\bibinfo{author}{\bibfnamefont{R.}~\bibnamefont{Kosloff}},
  \bibinfo{journal}{Annu. Rev. Phys. Chem.} \textbf{\bibinfo{volume}{45}},
  \bibinfo{pages}{145} (\bibinfo{year}{1994}).

\bibitem[{\citenamefont{Kosloff}(1988)}]{Kosloff88}
\bibinfo{author}{\bibfnamefont{R.}~\bibnamefont{Kosloff}}, \bibinfo{journal}{J.
  Phys. Chem.} \textbf{\bibinfo{volume}{92}}, \bibinfo{pages}{2087}
  (\bibinfo{year}{1988}).

\bibitem[{\citenamefont{Vibok and Balint-Kurti}(1992)}]{Vibok92a}
\bibinfo{author}{\bibfnamefont{A.}~\bibnamefont{Vibok}} \bibnamefont{and}
  \bibinfo{author}{\bibfnamefont{G.~G.} \bibnamefont{Balint-Kurti}},
  \bibinfo{journal}{J. Chem. Phys.} \textbf{\bibinfo{volume}{96}},
  \bibinfo{pages}{7615} (\bibinfo{year}{1992}).

\bibitem[{\citenamefont{Brumer and Shapiro}(1992)}]{PBrumer92}
\bibinfo{author}{\bibfnamefont{P.}~\bibnamefont{Brumer}} \bibnamefont{and}
  \bibinfo{author}{\bibfnamefont{M.}~\bibnamefont{Shapiro}},
  \bibinfo{journal}{Annu. Rev. Phys. Chem.} \textbf{\bibinfo{volume}{43}},
  \bibinfo{pages}{257} (\bibinfo{year}{1992}).

\bibitem[{\citenamefont{Brumer and Shapiro}(1986)}]{PBrumer86}
\bibinfo{author}{\bibfnamefont{P.}~\bibnamefont{Brumer}} \bibnamefont{and}
  \bibinfo{author}{\bibfnamefont{M.}~\bibnamefont{Shapiro}},
  \bibinfo{journal}{Chem. Phys. Lett.} \textbf{\bibinfo{volume}{126}},
  \bibinfo{pages}{541} (\bibinfo{year}{1986}).

\bibitem[{\citenamefont{Weiner et~al.}(1988)\citenamefont{Weiner, Heritage, and
  Kirschner}}]{AMWeiner88}
\bibinfo{author}{\bibfnamefont{A.~M.} \bibnamefont{Weiner}},
  \bibinfo{author}{\bibfnamefont{J.~P.} \bibnamefont{Heritage}},
  \bibnamefont{and} \bibinfo{author}{\bibfnamefont{E.~M.}
  \bibnamefont{Kirschner}}, \bibinfo{journal}{J. Opt. Soc. Am. B}
  \textbf{\bibinfo{volume}{5}}, \bibinfo{pages}{1563} (\bibinfo{year}{1988}).

\bibitem[{\citenamefont{Nelson et~al.}(1991)\citenamefont{Nelson, Weiner,
  Leaird, and Wiederrecht}}]{KANelson91}
\bibinfo{author}{\bibfnamefont{K.~A.} \bibnamefont{Nelson}},
  \bibinfo{author}{\bibfnamefont{A.~M.} \bibnamefont{Weiner}},
  \bibinfo{author}{\bibfnamefont{D.~E.} \bibnamefont{Leaird}},
  \bibnamefont{and} \bibinfo{author}{\bibfnamefont{G.~P.}
  \bibnamefont{Wiederrecht}}, \bibinfo{journal}{J. Opt. Soc. Am. B}
  \textbf{\bibinfo{volume}{8}}, \bibinfo{pages}{1264} (\bibinfo{year}{1991}).

\bibitem[{\citenamefont{Hammerich et~al.}(1989)\citenamefont{Hammerich, Muga,
  and Kosloff}}]{ADHammerich89}
\bibinfo{author}{\bibfnamefont{A.~D.} \bibnamefont{Hammerich}},
  \bibinfo{author}{\bibfnamefont{J.~G.} \bibnamefont{Muga}}, \bibnamefont{and}
  \bibinfo{author}{\bibfnamefont{R.}~\bibnamefont{Kosloff}},
  \bibinfo{journal}{Isr. J. Chem.} \textbf{\bibinfo{volume}{29}},
  \bibinfo{pages}{461} (\bibinfo{year}{1989}).

\end{thebibliography}
\end{document}